\documentclass[12pt]{iopart}

\usepackage{graphicx}
\usepackage{hyperref}
\usepackage{epstopdf}
\usepackage{amsthm}
\newtheorem{lemma}{\it Lemma}
\usepackage{array}
\usepackage{bm,bbm}
\usepackage{amssymb}
\usepackage{wrapfig}
\usepackage{caption}
\usepackage{subfigure}
\renewcommand{\thesubfigure}{\thefigure-\arabic{subfigure}}
\makeatletter
\renewcommand{\@thesubfigure}{(\thesubfigure)\space}
\renewcommand{\p@subfigure}{}
      \makeatother
\usepackage{iopams}
\begin{document}

\title{Finite-key analysis of a practical decoy-state high-dimensional quantum key distribution}
\author{Haize Bao$^{1,2}$,Wansu Bao$^{1,2,*}$,Yang Wang$^{1,2}$,Chun Zhou$^{1,2}$ and Ruike Chen$^{1,2}$}
\address{$^1$Zhengzhou Information Science and Technology Institute, Zhengzhou 450001, China}
\address{$^2$Synergetic Innovation Center of Quantum Information and Quantum Physics, University of Science and Technology of China, Hefei, Anhui 230026, China}
\ead{2010thzz@sina.com}

\vspace{10pt}
\begin{indented}
\item[]October 2015
\end{indented}

\begin{abstract}
  Compared with two-level quantum key distribution (QKD), high-dimensional QKD enable two distant parties to share a secret key at a higher rate. We provide a finite-key security analysis for the recently proposed practical high-dimensional decoy-state QKD protocol based on time-energy entanglement. We employ two methods to estimate the statistical fluctuation of the postselection probability and give a tighter bound on the secure-key capacity. By numerical evaluation, we show the finite-key effect on the secure-key capacity in different conditions. Moreover, our approach could be used to optimize parameters in practical implementations of high-dimensional QKD.
\end{abstract}

%
%
%
%
%
\section{Introduction}
Quantum key distribution (QKD) \cite{1,2,3} allows two authorized parties, Alice and Bob, to establish secret keys. However, the key generating rate is limited by various hardware constraints, such as the optical state generation and detection efficiency. To remedy this defect, a promising approach is high-dimensional QKD (HD-QKD). HD-QKD enables participants to maximize the secret-key capacity under technical constraints by encoding each photon with more than 1 bit of information. Moreover, compared with two-level QKD protocols, HD-QKD protocols are more robust to noise \cite{4}. HD-QKD encodes information in various photonic degrees of freedom, including position-momentum \cite{5}, time \cite{6,7,8,9}, time-energy \cite{10,11,12} and orbital angular momentum \cite{13,14}. So when photons are measured in such a high-dimensional Hilbert space, more than 1 bit of secure information can be shared by two parties per single-photon detected.

Recently, based on dispersive optics and time-energy entanglement, Jacob Mower \emph{et al.} \cite{15} proposed a high-dimensional QKD scenario, which is called dispersive optics QKD (DO-QKD) and proved its security against Gaussian collective attacks \cite{Furrer,Leverrier}. Instead of quantum bit error rate (QBER) in the BB84 QKD protocol, the security proof of DO-QKD uses the decrease in measurement correlations to estimate the amount of leaked information to the eavesdropper Eve. In DO-QKD, it is assumed that the photon pair produced by spontaneous parametric down conversion (SPDC) is an on-demand single-pair. However, because of the imperfection of the source \cite{16,17}, multi-pair emissions are always inevitable in real situation, causing the protocol is vulnerable to the photon number splitting (PNS) attack \cite{18,19,20} in lossy channels. In order to overcome the PNS attack, the decoy state method has been proposed \cite{21}. Recently, Zheshen Zhang \emph{et al.} \cite{22} extended the decoy state method to the DO-QKD scenario for an infinite number of decoy states, which is impractical.
Instead of an infinite number of decoy states, Darius Bunandar \emph{et al.} \cite{23} presented a practical decoy-state HD-QKD scheme and analyzed its security and feasibility. The analysis of decoy-state HD-QKD showed that under realistic experimental parameters, the security of HD-QKD with only two decoy states can approach the protocol with an infinite number of decoy states. In other words, the HD-QKD protocol with two decoy states is practical and feasible. The resource to establish secure keys is finite in practical implementations of QKD and many studies about the effect of finite-key have been finished \cite{Bratzik,Hayashi,Wang,Zhou}. So, how finite resources effect the security of the decoy-state HD-QKD protocol is worth further studying.

Catherine Lee \emph{et al.} \cite{24} have analyzed the finite-key effect on the security of the HD-QKD protocol without decoy states. In this paper, we analyze the security of a practical decoy-state HD-QKD protocol with finite resources. We consider statistical fluctuations in parameter estimation step in the finite-key regime. We employ two statistical methods to estimate the statistical fluctuation in Alice's and Bob's measurements and derive tight finite-key security bounds. By numerical evaluation, we evaluate the finite-key effect on the performance of the decoy-state HD-QKD with realistic constraints.

The paper is organized as follow: In Sec. 2, we briefly describe the decoy-state HD-QKD protocol. Sec. 3 introduces the formulas and two methods that are used in our finite-key analysis. Sec. 4 presents the fluctuation range and derives a tight security bound on the secure-key capacity with finite resources. The results of numerical evaluation are presented In Sec. 5. The conclusion is summarized in Sec. 6.
\begin{figure}
  \centering
  \includegraphics[width=0.8\textwidth]{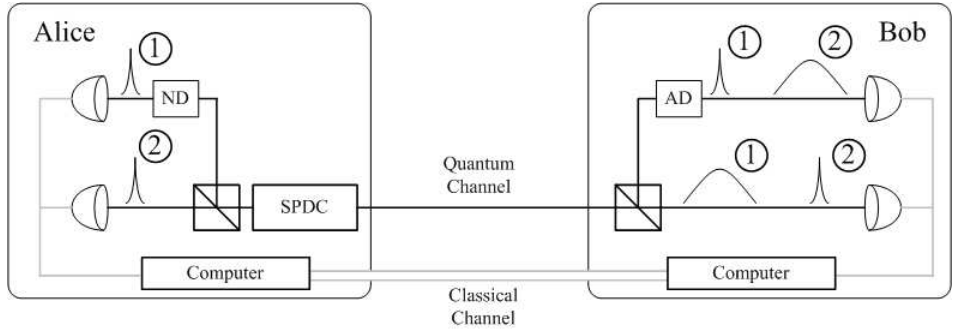}
  \caption{(Color online) Schematic of the HD-QKD setup. Alice holds the source, keeps one of a pair of photons produced by SPDC and sends the other to Bob. In case 1, Alice measures in the dispersed-time basis. In case 2, Alice measures in arrival-time basis. When Bob measures in the same basis as Alice, their measurements are correlated. ND is normal dispersion and AD is anomalous dispersion.}
  \label{1}
\end{figure}

\section{Protocol description}
The time-energy entanglement-based HD-QKD protocol that we consider is illustrated in Figure~\ref{1}. We assume that setups are completely controlled by Alice and Bob respectively. The quantum channel and the classical channel are authenticated. In our security analysis, for simplicity, Alice and Bob generate raw keys and estimate parameters in two bases separately \cite{25}.In particular, we extract the secret key from the coincidence that Alice and Bob both detect photons in arrival-time basis ($\mathbb{T}$ basis) and estimate parameters in dispersive-time basis ($\mathbb{D}$ basis). In addition, the source intensity is selected randomly in one of the three intensities $\mu $, ${{v}_{1}}$ and ${{v}_{2}}$, satisfying $\mu >{{v}_{1}}+{{v}_{2}}$ and ${{v}_{1}}>{{v}_{2}}\ge 0$. The protocol is descripted as follows:

\begin{enumerate}
  \item \textbf{State preparation:} Alice chooses independently an intensity $\lambda \in \left\{ \mu ,{{v}_{1}},{{v}_{2}} | \mu >{{v}_{1}}+{{v}_{2}} , {{v}_{1}}>{{v}_{2}}\ge 0 \right\}$ with probabilities ${{p}_{\mu }},{{p}_{{{v}_{1}}}}$ and $1-{{p}_{\mu }}-{{p}_{{{v}_{1}}}}$ and generates a biphoton state by SPDC. She keeps one of the biphoton and sends the other to Bob via a quantum channel.
  \item \textbf{State measurement:} Alice and Bob select a basis at  \{$\mathbb{T}$, $\mathbb{D}$\}. Both select $\mathbb{T}$ basis with the same probability ${{p}_{T}}$ and select $\mathbb{D}$ basis with the probability ${1-{p}_{T}}$. Then they perform measurements independently. After their measurements, Alice and Bob record their own outcomes respectively.
  \item \textbf{Classical information post-processing:} After all $N$ laser pulses are transmitted, Alice and Bob publish their basis and intensity choices over an authenticated public channel. To perform parameter estimation, Alice and Bob also announce their outcomes in $\mathbb{D}$ basis. They establish their distilled keys only from correlated timing events acquired in $\mathbb{T}$ basis. Note that we define $N$ as the number of laser pulses sent by Alice to Bob. After measuring postselection probabilities and multipliers, Alice and Bob can calculate Eve's Holevo information and then detect Eve's influence so that they can determine their information advantage over Eve. If the advantage is much greater than zero, they then apply error correction and privacy amplification on their data. As a result, some amount of secret key can be established.
\end{enumerate}

\section{Formulas and methods}
First, we shall introduce the postselection probability in general decoy-state QKD protocols, which is the probability of Alice and Bob registering at least one detection (due to a photon or a dark count) in a single measurement frame \cite{23}.

The entangled photon-pair from the SPDC source exhibits approximate Poisson distribution \cite{16,28}. Suppose the intensity of Alice's SPDC source is $\lambda $, in a single measurement frame, the probability ${{\Pr }_{n}}$ of generating $n$-photon pairs is
{\setlength\abovedisplayskip{0pt}
\setlength\belowdisplayskip{0pt}
\begin{center}
\begin{equation}
{{\Pr }_{n}}={\frac{{{\lambda }^{n}}}{n!}}{{e}^{-\lambda }}.
\end{equation}
\end{center}

Then the postselection probability can be written as
\begin{center}
\begin{equation}
{{P}_{\lambda }}=\displaystyle\sum\limits_{n=0}^{\infty }{{{\Pr }_{n}}{{\gamma }_{n}}}=\sum\limits_{n=0}^{\infty }{\frac{{{\lambda }^{n}}}{n!}}{{e}^{-\lambda }}{{\gamma }_{n}},
\end{equation}
\end{center}
where, ${{\gamma }_{n}}$ is the conditional probability of measuring at least one detection given $n$-photon pairs are emitted. With Eve's absence, we have \cite{23}
\begin{center}
\begin{equation}
{{\gamma }_{n}}=\left[ 1-{{\left( 1-{{\eta }_{Alice}} \right)}^{n}}\left( 1-{{p}_{d}} \right) \right]\left[ 1-{{\left( 1-{{\eta }_{Bob}}{{\eta }_{T}} \right)}^{n}}\left( 1-{{p}_{d}} \right) \right].
\end{equation}
\end{center}
Here, ${{\eta }_{Alice}}$ and ${{\eta }_{Bob}}$ are Alice and Bob's detector efficiencies respectively. ${{\eta }_{T}}$ is the transmittance of the quantum channel and ${{p}_{d}}$ is the dark count rate in a single measurement frame. In principle, Eve has the ability to affect the value of ${{\gamma }_{n}}$. Decoy states are utilized to estimate the ${{\gamma }_{n}}$ values from the postselection probabilities of different choices of $\lambda $ and then Eve's effect on the security can be estimated.

We consider $\varepsilon $-security \cite{29,30}, where $\varepsilon $ corresponds to the failure probability of the entire protocol. $\varepsilon $ is limited by other small parameters and has
\begin{center}
\begin{equation}
\varepsilon ={{\varepsilon }_{PE}}+{{\varepsilon }_{EC}}+\bar{\varepsilon }+{{\varepsilon }_{PA}}.
\end{equation}
\end{center}
All parameters ${{\varepsilon }_{PE}}$, ${{\varepsilon }_{EC}}$, $\bar{\varepsilon }$ and ${{\varepsilon }_{PA}}$ can independently be fixed at arbitrarily low values:
\begin{enumerate}
  \item ${{\varepsilon }_{PE}}$ is the failure probability of parameter estimation procedure and can be made as low as desired simply by increasing the size of the sample used for parameter estimation.
  \item ${{\varepsilon }_{EC}}$ is the failure probability of error correction procedure and can be decreased by computing a hash of Alice's and Bob's bit strings after the reconciliation respectively and comparing it publicly.
  \item $\bar{\varepsilon }$ is a smooth min-entropy parameter and ${{\varepsilon }_{PA}}$ is the failure probability of privacy amplification procedure. Both of them are virtual parameters and can be optimized in the computation.
\end{enumerate}
So we can chose security parameter to an arbitrarily small value.

The finite-size secure-key capacity for the HD-QKD protocol can be rewritten as \cite{29,30}:
\begin{center}
\begin{equation}
\Delta I={{R}_{HD}}-\frac{1}{{{p}_{\mu }}p_{T}^{2}N}{{\log }_{2}}\frac{2}{{{\varepsilon }_{EC}}}-\frac{2}{{{p}_{\mu }}p_{T}^{2}N}{{\log }_{2}}\frac{1}{{{\varepsilon }_{PA}}}-\left(2d+3\right)\sqrt{\frac{1}{{{p}_{\mu }}p_{T}^{2}N}{{\log }_{2}}\frac{2}{{\bar{\varepsilon }}}}.
\label{deltaI}
\end{equation}
\end{center}
Zhang \cite{22} has given out the ${{R}_{HD}}$ expression in the decoy-state HD-QKD protocol:
\begin{equation}
\setlength{\abovedisplayskip}{6pt}
{{R}_{HD}}=\beta I\left( A;B \right)-(1-{{K}_{\mu }}){{I}_{R}}-{{K}_{\mu }}\phi _{{{\zeta }_{t}},{{\zeta }_{\omega }}}^{UB}\left( A;E \right).
\setlength{\belowdisplayskip}{6pt}
\label{RHD}
\end{equation}
Here, as it's described in Ref\cite{22,23}, $\beta$ is the reconciliation efficiency and ${{K}_{\mu }}=\mu {{e}^{-\mu }}{{\gamma }_{1}}/{{P}_{\mu }}$ is the fraction of postselected events. ${{I}_{R}}$ is the number of random bits shared between Alice and Bob, which is $log_2 d$ here. $I\left( A;B \right)$ is the mutual information between Alice and Bob. $\phi _{{{\zeta }_{t}},{{\zeta }_{\omega }}}^{UB}\left( A;E \right)$ is an upper bound on Eve's Holevo information under Gaussian collective attacks given the excess-noise factor ${{\zeta }_{t}}$ and ${{\zeta }_{\omega }}$ for the timing and the frequency correlations, respectively.

Then combing Eq.~\ref{deltaI} and Eq.~\ref{RHD}, we obtain the finite-size secure-key capacity for the decoy-state HD-QKD protocol:
\begin{eqnarray}
  \Delta I=&\beta I\left( A;B \right)-(1-{{K}_{\mu }}){{I}_{R}}-{{K}_{\mu }}\phi _{{{\zeta }_{t}},{{\zeta }_{\omega }}}^{UB}\left( A;E \right)-\frac{1}{{{p}_{\mu }}p_{T}^{2}N}{{\log }_{2}}\frac{2}{{{\varepsilon }_{EC}}}\nonumber\\
   &-\frac{2}{{{p}_{\mu }}p_{T}^{2}N}{{\log }_{2}}\frac{1}{{{\varepsilon }_{PA}}}-\left( 2d+3 \right)\sqrt{\frac{1}{{{p}_{\mu }}p_{T}^{2}N}{{\log }_{2}}\frac{2}{{\bar{\varepsilon }}}}
\end{eqnarray}

\section{Parameter estimation in finite-key}

We extend the decoy-state analysis proposed in Ref. \cite{23} to the case of finite-key. In the two decoy-state condition, the value of ${{\gamma }_{0}}$ and ${{\gamma }_{1}}$ satisfy
\begin{equation}
\left\{ \begin{array}{l}
   \gamma _{0}^{LB,\left\{ {{v}_{1}},{{v}_{2}} \right\}}=\max \left\{ \frac{{{v}_{1}}P_{{{v}_{2}}}^{-}{{e}^{{{v}_{2}}}}-{{v}_{2}}P_{{{v}_{1}}}^{+}{{e}^{{{v}_{1}}}}}{{{v}_{1}}-{{v}_{2}}},p_{d}^{2} \right\}\\\\
   \gamma _{0}^{UB,\left\{ {{v}_{1}},{{v}_{2}} \right\}}={{p}_{d}} \\\\
   \gamma _{0}^{LB,\left\{ {{v}_{1}},{{v}_{2}} \right\}}\le {{\gamma }_{0}}\le \gamma _{0}^{UB,\left\{ {{v}_{1}},{{v}_{2}} \right\}}
\end{array} \right.
\label{Gamma0}
\end{equation}
and
\begin{center}
\begin{equation}
\label{Gamma1}
{{\gamma }_{1}}\ge \frac{\mu }{\mu {{v}_{1}}-\mu {{v}_{2}}-v_{1}^{2}+v_{2}^{2}}\left[P_{{{v}_{1}}}^{-}{{e}^{{{v}_{1}}}}-P_{{{v}_{2}}}^{+}{{e}^{{{v}_{2}}}}-\frac{v_{1}^{2}-v_{2}^{2}}{{{\mu }^{2}}}\left( P_{\mu }^{+}{{e}^{\mu }}-{{\gamma }_{0}} \right) \right],
\end{equation}
\end{center}
where ${{v}_{1}}$ and ${{v}_{2}}$ are two different decoy-state intensities, and ${{p}_{d}}$ is dark count rate. Besides, ${{P}_{\lambda }}$ is the measured result and satisfies the relationship:
\begin{equation}
P_{\lambda }^{\pm }={{P}_{\lambda }}\pm \Delta \left( N,{{\varepsilon }_{PE}} \right).
\label{Pf}
\end{equation}
Here, $\Delta \left( N,{{\varepsilon }_{PE}} \right)$ is the fluctuation range of the postselection ${{P}_{\lambda }}$ which is related to $N$ and ${{\varepsilon }_{PE}}$. We will show the accurate expression of $\Delta$ in Eq.~\ref{deltaH} and Eq.~\ref{deltaC} in Sec. 4 with different approaches.

By combining Eq.~\ref{Gamma0}, Eq.~\ref{Gamma1} and Eq.~\ref{Pf}, we can get a lower bound on ${{K}_{\mu }}$:
\begin{center}
\begin{eqnarray}
     {{K}_{\mu }}&=\frac{\mu {{e}^{-\mu }}{{\gamma }_{1}}}{{{P}_{\mu }}}\ge \frac{\mu {{e}^{-\mu }}{{\gamma }_{1}}}{P_{\mu }^{+}}=\frac{{{\mu }^{2}}}{\mu {{v}_{1}}-\mu {{v}_{2}}-v_{1}^{2}+v_{2}^{2}}\nonumber\\
     &\times\left[ \frac{P_{{{v}_{1}}}^{-}}{P_{\mu }^{+}}{{e}^{{{v}_{1}}-\mu }}-\frac{P_{{{v}_{2}}}^{+}}{P_{\mu }^{+}}{{e}^{{{v}_{2}}-\mu }}-\frac{v_{1}^{2}-v_{2}^{2}}{{{\mu }^{2}}}\left( 1-\frac{\gamma _{0}^{LB,\left\{ {{v}_{1}},{{v}_{2}} \right\}}{{e}^{-\mu }}}{P_{\mu }^{+}} \right) \right].
 \label{Kmiu1}
\end{eqnarray}
\end{center}
In the other hand, we can obtain another lower bound on ${{K}_{\mu }}$ by using the postselection probability of one of two decoy states directly:
\begin{center}
\begin{equation}
   {{K}_{\mu }}=\frac{\mu {{e}^{-\mu }}{{\gamma }_{1}}}{{{P}_{\mu }}}\ge \frac{\mu {{e}^{-\mu }}{{\gamma }_{1}}}{P_{\mu }^{+}} =\frac{{{\mu }^{2}}}{\mu \lambda -{{\lambda }^{2}}}\left[ \frac{P_{\lambda }^{-}}{P_{\mu }^{+}}{{e}^{\lambda -\mu }}-\frac{{{\lambda }^{2}}}{{{\mu }^{2}}}-\frac{{{\mu }^{2}}-{{\lambda }^{2}}}{{{\mu }^{2}}}\frac{\gamma _{0}^{UB,\left\{ {{v}_{1}},{{v}_{2}} \right\}}{{e}^{-\mu }}}{P_{\mu }^{+}} \right].
   \label{Kmiu2}
\end{equation}
\end{center}
Here, $\lambda ={{v}_{1}}$ or ${{v}_{2}}$ if ${{v}_{2}}\ne 0$  or $\lambda ={{v}_{1}}$ if ${{v}_{2}}=0$.

Therefore, combining Eq.~\ref{Kmiu1} and Eq.~\ref{Kmiu2}, we get
\begin{center}
\begin{eqnarray}
&{{K}_{\mu }}\ge K_{\mu }^{LB,\left\{ {{v}_{1}},{{v}_{2}} \right\}}=\max \left\{ \frac{{{\mu }^{2}}}{\mu {{v}_{1}}-\mu {{v}_{2}}-v_{1}^{2}+v_{2}^{2}}\right.\nonumber\\
   &\left[ \frac{P_{{{v}_{1}}}^{-}}{P_{\mu }^{+}}{{e}^{{{v}_{1}}-\mu }}-\frac{P_{{{v}_{2}}}^{+}}{P_{\mu }^{+}}{{e}^{{{v}_{2}}-\mu }} \right.\left.-\frac{v_{1}^{2}-v_{2}^{2}}{{{\mu }^{2}}}\left( 1-\frac{\gamma _{0}^{LB,\left\{ {{v}_{1}},{{v}_{2}} \right\}}{{e}^{-\mu }}}{P_{\mu }^{+}} \right) \right], \\
   &\left. \frac{{{\mu }^{2}}}{\mu \lambda -{{\lambda }^{2}}}\left[ \frac{P_{\lambda }^{-}}{P_{\mu }^{+}}{{e}^{\lambda -\mu }}- \right.\frac{{{\lambda }^{2}}}{{{\mu }^{2}}}-\left. \frac{{{\mu }^{2}}-{{\lambda }^{2}}}{{{\mu }^{2}}}\frac{\gamma _{0}^{UB,\left\{ {{v}_{1}},{{v}_{2}} \right\}}{{e}^{-\mu }}}{P_{\mu }^{+}} \right] \right\}\nonumber.
\end{eqnarray}
\end{center}

The excess noise is the noise added by the channel beyond the fundamental shot noise and corresponds to information loss in the QKD system. In the HD-QKD protocol, timing and frequency excess noise, ${{\zeta }_{t}}$ and ${{\zeta }_{\omega }}$, are used to parameterize the decrease, caused by Eve's attack, in the correlations of Alice and Bob¡¯s measurements.  Their respective average multipliers ${{\Phi }_{t,\lambda }}$ and ${{\Phi }_{\omega ,\lambda }}$ could be measured by Alice and Bob, and have the following relationship \cite{23}
\begin{center}
\begin{equation}
{{\Phi }_{x,\lambda }}={{K}_{\lambda }}\left( 1+{{\zeta }_{x}} \right)+\left( 1-{{K}_{\lambda }} \right)\Delta {{\Phi }_{x}},
\end{equation}
\end{center}
for $x=t$ or $\omega $ and $\lambda \in \left\{ \mu ,{{v}_{1}},{{v}_{2}} \right\}$.

Similar to analysis in ${{K}_{\mu }}$ , the upper bound on excess noise can be obtained:
\begin{center}
\begin{eqnarray}
 &{{\zeta }_{x}}\le \zeta _{x}^{UB,\left\{ {{v}_{1}},{{v}_{2}} \right\}}=\min \left\{ \min \left\{ \frac{\mu {{e}^{-\mu }}}{\left( {{\lambda }_{1}}-{{\lambda }_{2}} \right)K_{\mu }^{LB,\left\{ {{v}_{1}},{{v}_{2}} \right\}}} \right. \right. \\
 &\times \left. \left( {{\Phi }_{x,{{\lambda }_{1}}}}\frac{P_{{{\lambda }_{1}}}^{+}}{P_{\mu }^{+}}{{e}^{{{\lambda }_{1}}}}-{{\Phi }_{x,{{\lambda }_{2}}}}\frac{P_{{{\lambda }_{2}}}^{-}}{P_{\mu }^{+}}{{e}^{{{\lambda }_{2}}}} \right) \right\},\left. \min \left\{ {{e}^{\lambda -\mu }}\frac{\mu P_{\lambda }^{+}}{\lambda P_{\mu }^{+}}\frac{{{\Phi }_{x,\lambda }}}{K_{\mu }^{LB,\left\{ {{v}_{1}},{{v}_{2}} \right\}}} \right\} \right\}-1, \nonumber
\end{eqnarray}
\end{center}
where, $x=t$ or $\omega $, $\lambda ,{{\lambda }_{1}},{{\lambda }_{2}}\in \left\{ \mu ,{{v}_{1}},{{v}_{2}} \right\}$ and ${{\lambda }_{1}}>{{\lambda }_{2}}$.
In the single decoy-state condition, bounds on ${{K}_{\mu }}$ and ${{\zeta }_{x,\lambda }}$ can be directly deduced:
\begin{center}
\begin{equation}
 {{K}_{\mu }}\ge K_{\mu }^{LB,\left\{ v \right\}}=\frac{{{\mu }^{2}}}{\mu v-{{v}^{2}}}\left[ \frac{P_{v}^{-}}{P_{\mu }^{+}}{{e}^{v-\mu }}- \right.\frac{{{v}^{2}}}{{{\mu }^{2}}}-\left. \frac{{{\mu }^{2}}-{{v}^{2}}}{{{\mu }^{2}}}\frac{\gamma _{0}^{UB,\left\{ v \right\}}{{e}^{-\mu }}}{P_{\mu }^{+}} \right],
\end{equation}
\end{center}
and
\begin{center}
\begin{eqnarray}
   {{\zeta }_{x}}\le \zeta _{x}^{UB,\left\{ v \right\}}=\min &\left\{ \frac{\mu }{\left( \mu -v \right)K_{\mu }^{LB,\left\{ v \right\}}}\left( {{\Phi }_{x,\mu }}-{{\Phi }_{x,v}}\frac{P_{v}^{-}}{P_{\mu }^{+}}{{e}^{v-\mu }} \right) \right., \nonumber\\
  &\left. {{\mathop{\min }}\atop{\scriptstyle\lambda \in \left\{ \mu ,v \right\}}}\,\left\{ {{e}^{\lambda -\mu }}\frac{\mu P_{\lambda }^{+}}{\lambda P_{\mu }^{+}}\frac{{{\Phi }_{x,\lambda }}}{K_{\mu }^{LB,\left\{ {{v}_{1}},{{v}_{2}} \right\}}} \right\} \right\}-1,
\end{eqnarray}
\end{center}
for $x=t$ or $\omega $.

In most of finite-key analysis in QKD schemes, for independent and random measurements, Hoeffding's inequality \cite{32} and Chernoff bound \cite{33} are always used in statistical fluctuation analysis \cite{34,35}. Therefore, we take Hoeffding's inequality and Chernoff bound respectively to deal with the statistical fluctuation.

Suppose the intensity of photons used in parameter estimation is $\lambda$, we denote the number of the photon as $N_{\lambda}$. In particular, $N_{\lambda}$ is expressed as $p_{\lambda}\left( 1-p_T \right) ^2N$ in the decoy-state HD-QKD.
\begin{lemma}\cite{32}
\label{lemma1}
Let $X_1$, $X_2$,...,$X_n$, be a set of independent Bernoulli random variables that satisfy $P\left( X_i=1 \right) =p_i$, and let $X=\sum_{i=1}^n{X_i}$ and $\alpha=E\left( X \right) =\sum_{i=1}^n{p_i}$, where $E\left( \cdot \right) $ denotes the mean value. Let $\beta$ be the observed outcome of $X$ for a given trial (i.e., $\beta\in\mathbb{N^+}$). For certain $\varepsilon_{H}, \hat{\varepsilon}_{H}>0$, we have that $\beta$ satisfies
\begin{equation}
\centering
\beta =\alpha +\theta_{H},
\end{equation}
except with error probability $\gamma =\varepsilon_{H} +\hat{\varepsilon}_{H}$, where the parameter $\theta_{H} \in \left[ -\tau_{H} ,\hat{\tau}_{H} \right] $, with $\tau_{H} =g\left( N ,{\varepsilon_{H}} \right)$, $\hat{\tau}_{H}=g\left( N ,\hat{\varepsilon}_{H}\right) $ and $g_H\left( x,y \right) =\sqrt{x/2\ln \left( y^{-1} \right)}$. Here $\varepsilon_{H} $($\hat{\varepsilon}_{H}$) denotes the probability that $\beta <\alpha -\tau_{H}$ ( $\beta >\alpha +\hat{\tau}_{H}$).
\end{lemma}
In particular, we suppose the intensity of photons used in parameter estimation is $\lambda$ and denote the number of the photon as $N_{\lambda}$. For simplifying analysis, we set $\varepsilon_{PE}={\varepsilon}_{H}+\hat{\varepsilon}_{H}$ and ${\varepsilon}_{H}=\hat{\varepsilon}_{H}$. According to Lemma ~\ref{lemma1}, we have
\begin{equation}
\centering
P_{\lambda}^{*}N_{\lambda}+\sqrt{N_{\lambda}/2\ln \left( 2\varepsilon ^{-1} \right)}\geq P_{\lambda}N_{\lambda}\geq P_{\lambda}^{*}N_{\lambda}-\sqrt{N_{\lambda}/2\ln \left( 2\varepsilon ^{-1} \right)},
\end{equation}
where $P_{\lambda}^{*}$ is the real value of the postselection probability.
In detail, $N_{\lambda}$ can be expressed as $p_{\lambda}\left( 1-p_T \right) ^2N$, so we obtain
\begin{equation}
\label{deltaH}
\Delta \left( N,{{\varepsilon }_{PE}} \right)={{\Delta }_{H}}=\sqrt{\frac{1}{2{{p}_{\lambda }}{{\left( 1-{{p}_{T}} \right)}^{2}}N}\ln \frac{2}{{{\varepsilon }_{PE}}}}
\end{equation}
and
\begin{equation}
\qquad P_{\lambda }^{\pm }={{P}_{\lambda }}\pm \sqrt{\frac{1}{2{{p}_{\lambda }}{{\left( 1-{{p}_{T}} \right)}^{2}}N}\ln \frac{2}{{{\varepsilon }_{PE}}}}
\end{equation}
In the other hand, we consider multiplicative Chernoff bound \cite{Alon,Angluin}.
\begin{lemma}\cite{Alon}
\label{lemma2}
Let $X_1$, $X_2$,...,$X_n$, be a set of independent Bernoulli random variables that satisfy $P\left( X_i=1 \right) =p_i$, and let $X=\sum_{i=1}^n{X_i}$ and $\alpha=E\left( X \right) =\sum_{i=1}^n{p_i}$, where $E\left( \cdot \right) $ denotes the mean value. Let $\beta$ be the observed outcome of $X$ for a given trial (i.e., $\beta\in\mathbb{N^+}$) and $\alpha_L=\beta-\sqrt{{N/2}\ln \epsilon ^{-1}}$ for certain $\epsilon>0$. When $\left( 2\varepsilon_C ^{-1} \right)^{{1}/{\alpha_L}}\le e^{\left[3/4\sqrt{2} \right] ^2}$ and $\left( \hat{\varepsilon}_C^{-1} \right) ^{{1}/{\alpha_L}}<e^{\frac{1}{3}}$ for certain $\varepsilon_C ,\hat{\varepsilon}_C>0$, we have that $\beta$ satisfies
\begin{equation}
\centering
\qquad\qquad\qquad\qquad\beta =\alpha +\theta_{C},
\end{equation}
except with error probability $\gamma =\epsilon+\varepsilon_C +\hat{\epsilon}_C$, where the parameter $\theta \in \left[ -\tau_{C} ,\hat{\tau}_{C} \right] $, with $\tau_C=g\left( \beta ,{\varepsilon_C ^4}/{16} \right)$, $\hat{\tau}_C=g\left( \beta ,\varepsilon_C ^{3/2} \right) $ and $g\left( x,y \right) =\sqrt{2x\ln \left( y^{-1} \right)}$. Here $\varepsilon_C $($\hat{\varepsilon}_C$) denotes the probability that $\beta <\alpha -\tau_C$ ( $\beta >\alpha +\hat{\tau}_C$).
\end{lemma}
When the amount of signals is great enough, in Lemma~\ref{lemma2}, the condition, $\left( 2\varepsilon_C ^{-1} \right)^{{1}/{\alpha_L}}\le e^{\left[3/4\sqrt{2} \right] ^2}$ and $\left( \hat{\varepsilon}_C^{-1} \right) ^{{1}/{\alpha_L}}<e^{\frac{1}{3}}$ for certain $\varepsilon_C ,\hat{\varepsilon}_C>0$, is met easily. Similar to the above analysis, we set $\varepsilon_{PE}={\epsilon}+{\varepsilon}_{C}+\hat{\varepsilon}_{C}$ and ${\epsilon}={\varepsilon}_{H}=\hat{\varepsilon}_{H}$. Then we have
\begin{equation}
\label{deltaC}
\left\{ \begin{array}{l}
\Delta \left( N,\varepsilon _{PE} \right) ^{+}=\sqrt{\frac{2P_{\lambda}}{p_{\lambda}\left( 1-p_T \right) ^2N}\ln \frac{16}{\left({\varepsilon _{PE}}/{3}\right)^{4}}} \\\\
\Delta \left( N,\varepsilon _{PE} \right) ^{-}=\sqrt{\frac{2P_{\lambda}}{p_{\lambda}\left( 1-p_T \right) ^2N}\ln \frac{1}{\left({\varepsilon _{PE}}/{3}\right)^{3/2}}}
\end{array} \right.
\end{equation}
and
\begin{equation}
\left\{ \begin{array}{l}
{P}^{+}={P}_{\lambda}+\sqrt{\frac{2{P}_{\lambda}}{{p}_{\lambda}\left( 1-{p}_{T} \right) ^2{N}}\ln \frac{16}{\left( \varepsilon _{PE}/{3} \right) ^4}}\\\\
{P}^{-}={P}_{\lambda}-\sqrt{\frac{2{P}_{\lambda}}{{p}_{\lambda}\left( 1-{p}_{T} \right) ^2{N}}\ln \frac{1}{\left( \varepsilon _{PE}/{3} \right) ^{3/2}}}
\end{array} \right.
\end{equation}
After above discussion, we can make numerical evaluation for all case of the finite-key practical decoy-state HD-QKD protocol analyzed in this paper.

\section{Numerical evaluation}
Here, we consider a fiber-based QKD system model that uses the recent decoy-state HD-QKD protocol experimental parameters \cite{23,36}: propagation loss $\alpha =$0.2 dB/km; detector timing jitter ${{\delta }_{j}}=20$ ps; dark count rate ${{R}_{dc}}=1000 {{s}^{-1}}$; reconciliation efficiency $\beta =0.9$; ${{I}_{R}}={{\log }_{2}}d$. Then we have the transmittance ${{\eta }_{T}}={{10}^{-\alpha L/10}}$, where $L$ is the length of the quantum channel in km. Besides, we also assume that Alice and Bob have the same detector efficiencies: ${{\eta }_{Alice}}={{\eta }_{Bob}}=0.93$ \cite{37}.

Our purpose is to test whether the protocol can be utilized to distill a secret key for some given conditions, so, from this point of view, factors, ${{p}_{T}}$ and ${{p}_{\lambda }}$, have little effect on the final rate \cite{30}. In particular, we let ${{p}_{T}}=50\%$ and ${{p}_{\mu }}:{{p}_{{{v}_{1}}}}:{{p}_{{{v}_{2}}}}=7:2:1$ in the two decoy-state condition while ${{p}_{\mu }}:{{p}_{v}}=4:1$ in the one decoy-state condition.

We assume ${{\varepsilon }_{PE}}={{\varepsilon }_{EC}}=\bar{\varepsilon }={{\varepsilon }_{PA}}={{10}^{-10}}$ \cite{38}. We take coherent time ${{\delta }_{coh}}=30\ $ps or all $d$ values and correlation time ${{\delta }_{cor}}=d{{\delta }_{coh}}$, which is easily controlled \cite{36}. The measurement frame duration ${{T}_{f}}$ is chosen to be ${{T}_{f}}=2\sqrt{2\ln 2}{{\delta }_{coh}}$ \cite{22}. For simplicity, we also assume two excess-noise factors are equal, ${{\zeta }_{t}}={{\zeta }_{\omega }}=\zeta $. The change in correlation time due to Eve's interaction can be assumed to be ${{\delta }_{\Delta }}=\left( \sqrt{1+\zeta }-1 \right){{\delta }_{cor}}=10\ ps$ \cite{24}, which doesn't lose generality.

Based on the Hoeffding's inequality, Figure~\ref{fig2} plots the secure-key capacity of decoy-state HD-QKD in the case of the Schmidt number $d=8$ and $\mu$= 0.01, 0.10 and 0.25. Figure~\ref{fig:2-1}, ~\ref{fig:2-2} and ~\ref{fig:2-3}  plot in the two-decoy-state HD-QKD protocol and Figure ~\ref{fig:2-4}, ~\ref{fig:2-5} and ~\ref{fig:2-6} plot in the single-decoy-state HD-QKD protocol.
\begin{figure}
  \centering
  \subfigure[]{\label{fig:2-1}\includegraphics[width=2.0in]{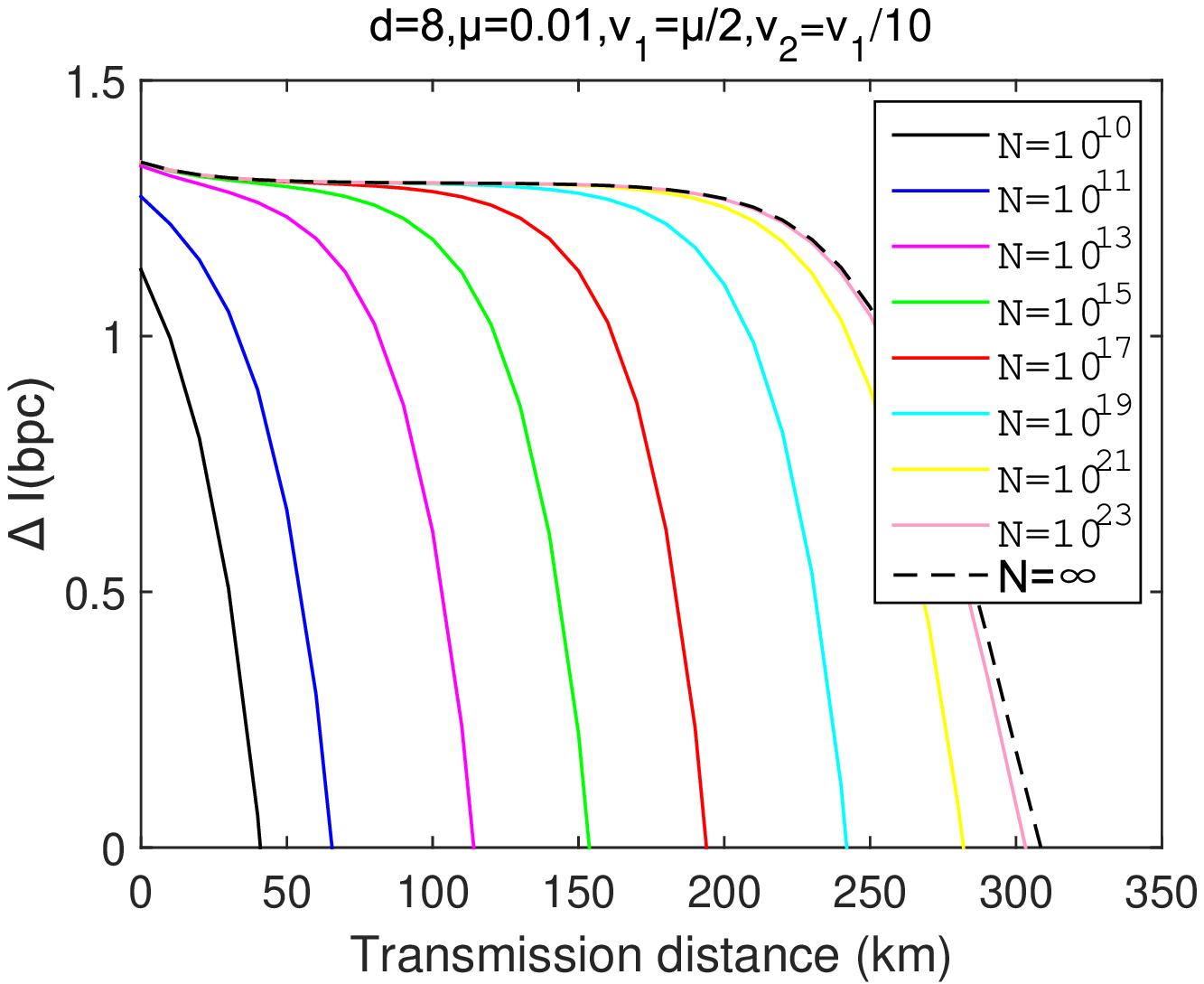}}
  \subfigure[]{\label{fig:2-2}\includegraphics[width=2.0in]{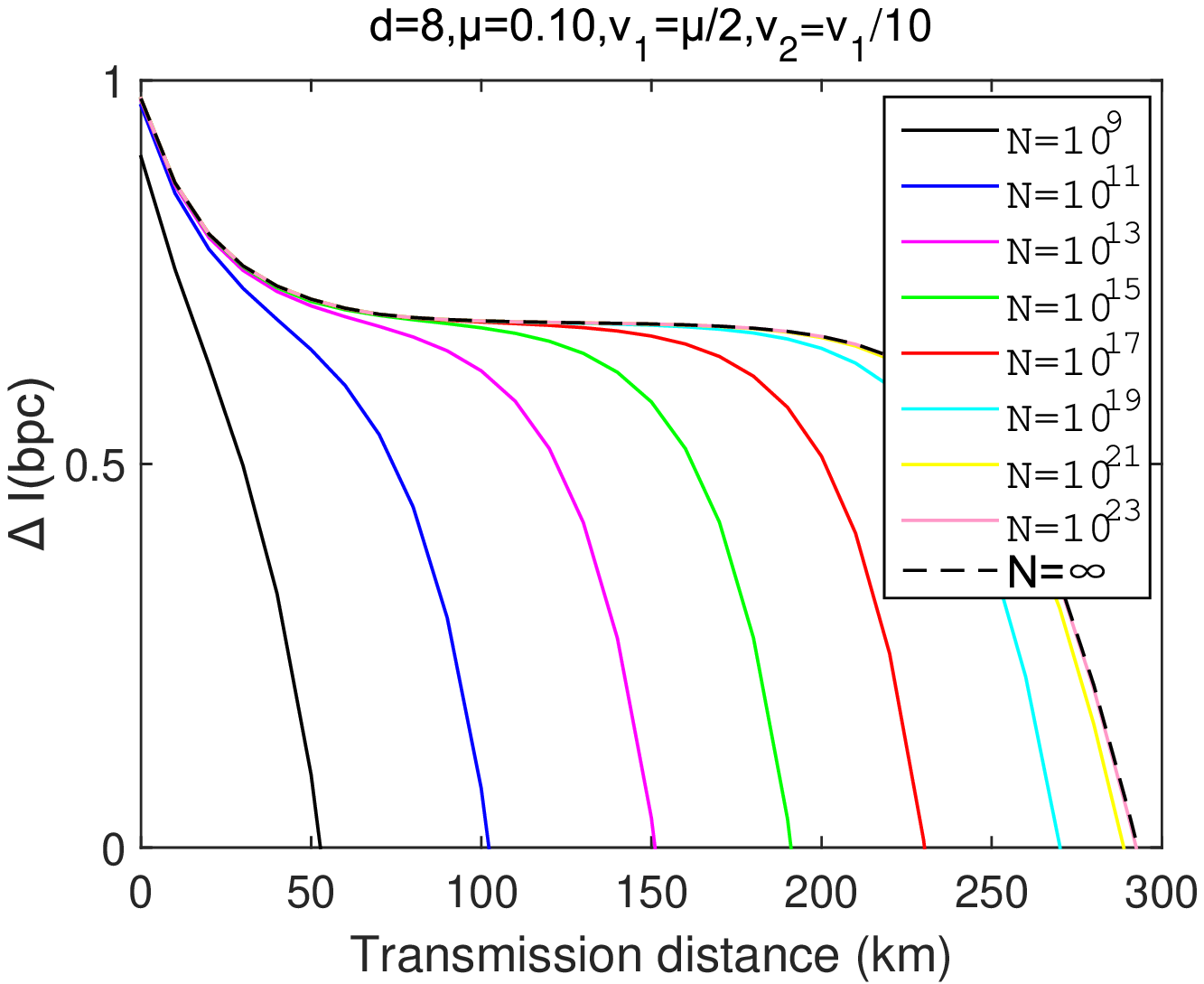}}
  \subfigure[]{\label{fig:2-3}\includegraphics[width=2.0in]{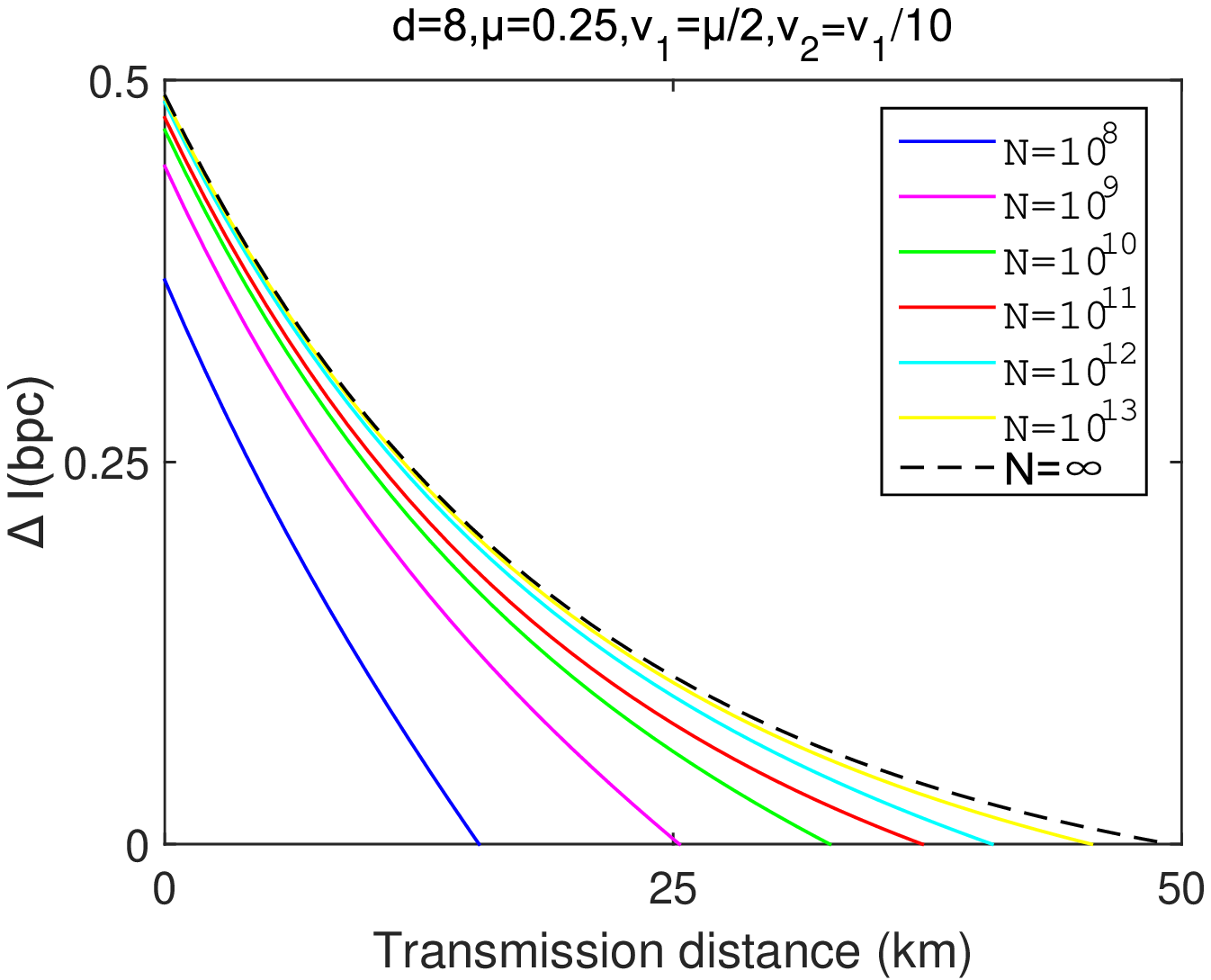}}
  \\
  \subfigure[]{\label{fig:2-4}\includegraphics[width=2.0in]{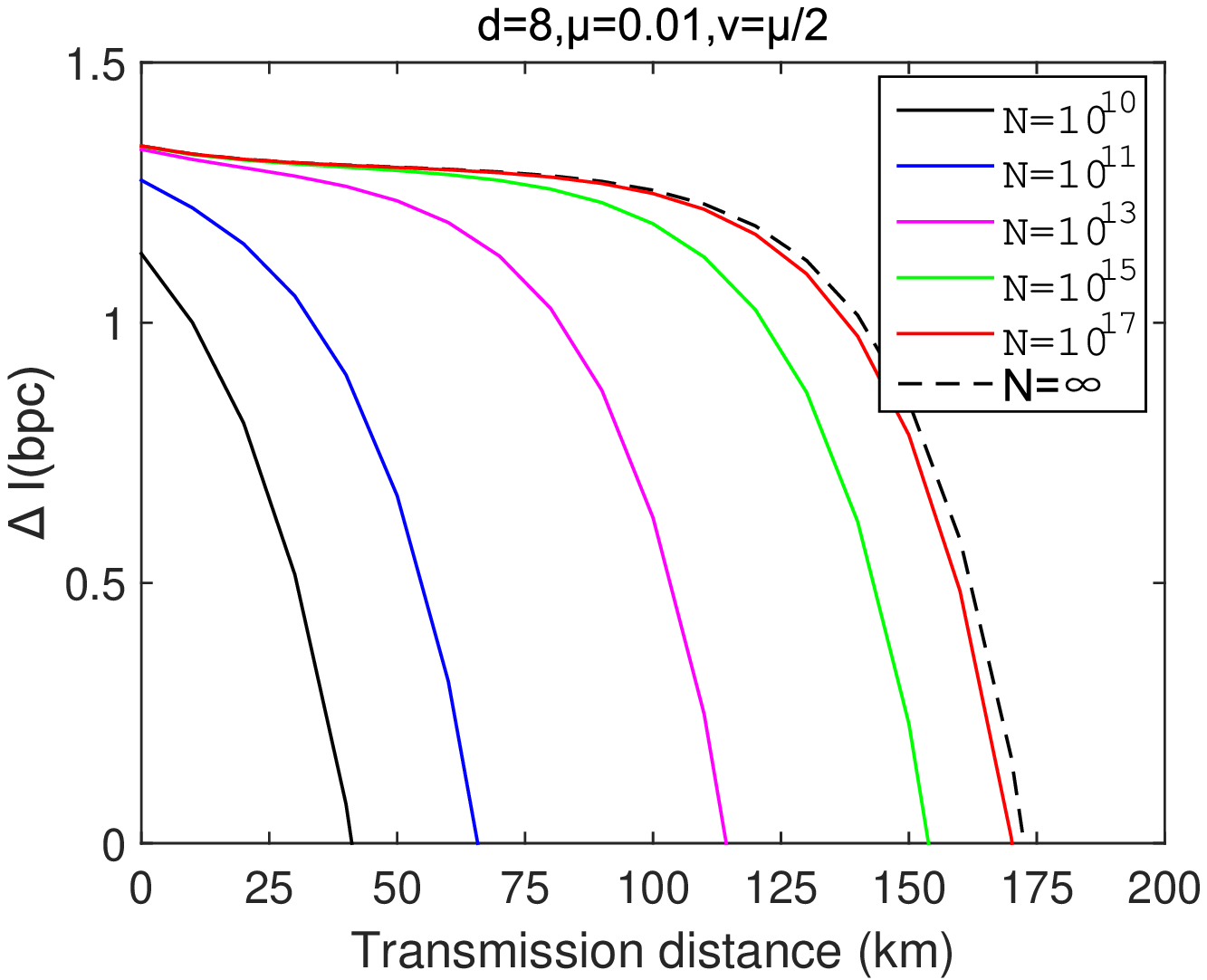}}
  \subfigure[]{\label{fig:2-5}\includegraphics[width=2.0in]{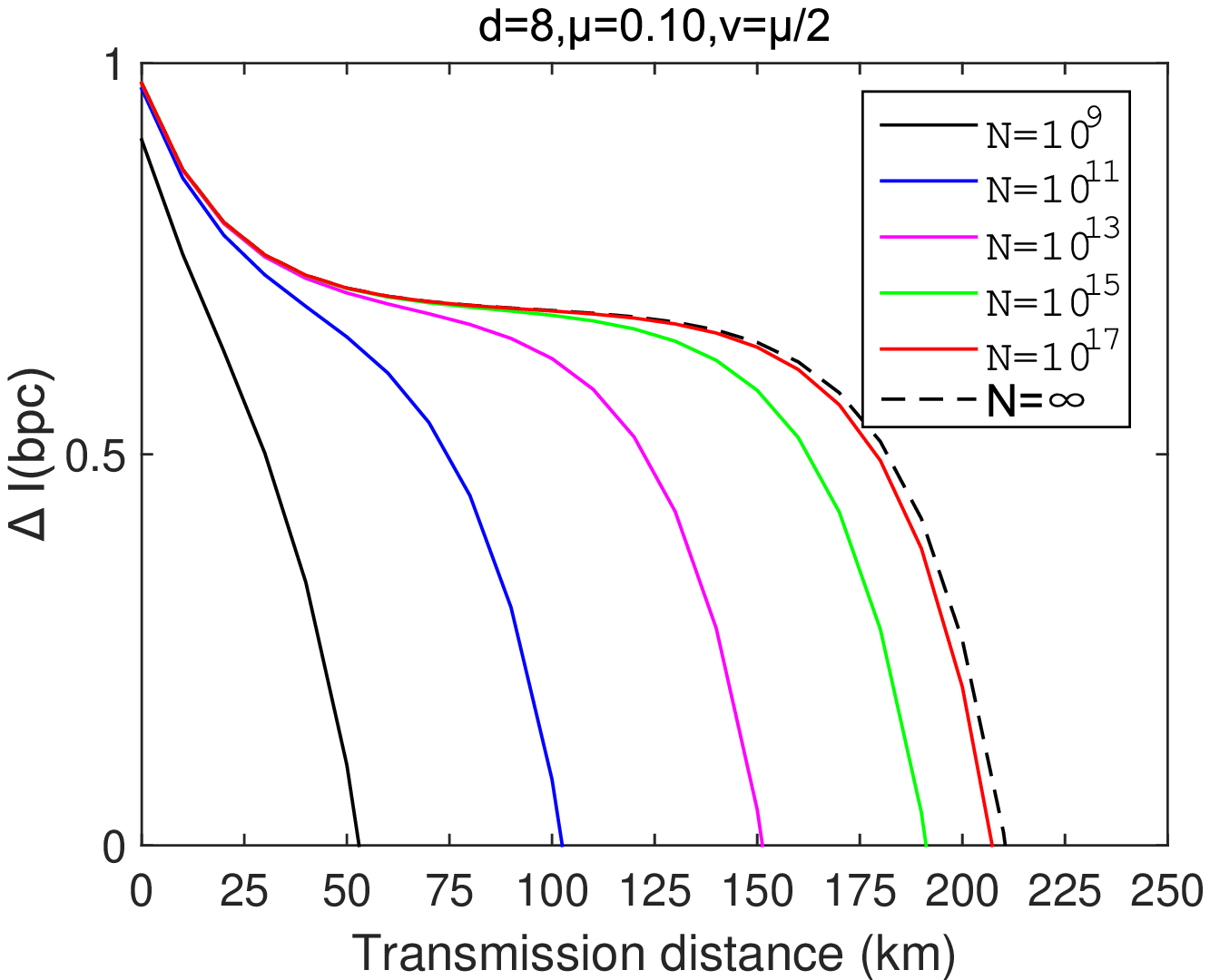}}
  \subfigure[]{\label{fig:2-6}\includegraphics[width=2.0in]{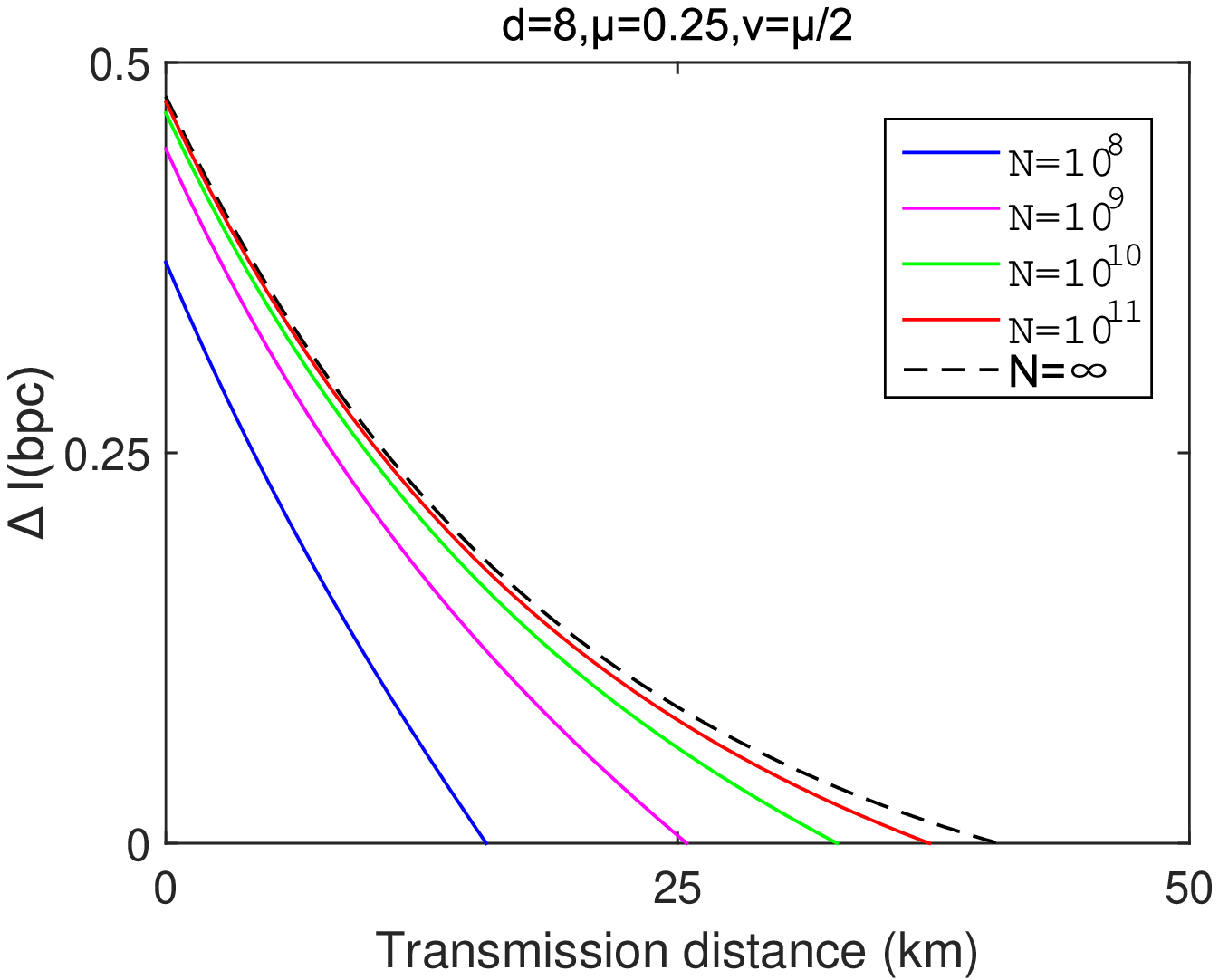}}
  \caption{(Color online) Finite-size secure-key capacities in bpc versus transmission distance based on Hoeffding's inequality. Top panels ( (2-1)(2-2)(2-3) ) show the two-decoy-state case, where ${{v}_{1}}={\mu }/{2}\;$, ${{v}_{2}}={{{v}_{1}}}/{10}\;$. Bottom panels ( (2-4)(2-5)(2-6) ) show the case single-decoy-state, where $v={\mu }/{2}\;$. Solid lines show finite-size secure-key capacities with different $N$, which is the number of transmission signals. From left to right, $N$ gets bigger gradually. The dash line describe the trend of the infinite-size secure-key capacities.}
  \label{fig2}
\end{figure}

To compare effects of Hoeffding's inequality and Chernoff bound in parameter estimation, based on Chernoff bound, Figure~\ref{fig3} plots the secure-key capacity of decoy-state HD-QKD in the case of the same condition as Figure~\ref{fig2}. Figure~\ref{fig:3-1}, ~\ref{fig:3-2} and ~\ref{fig:3-3}  plot in the two-decoy-state HD-QKD protocol and Figure ~\ref{fig:3-4}, ~\ref{fig:3-5} and ~\ref{fig:3-6} plot in the single-decoy-state HD-QKD protocol.
\begin{figure}
  \centering
  \subfigure[]{\label{fig:3-1}\includegraphics[width=2.0in]{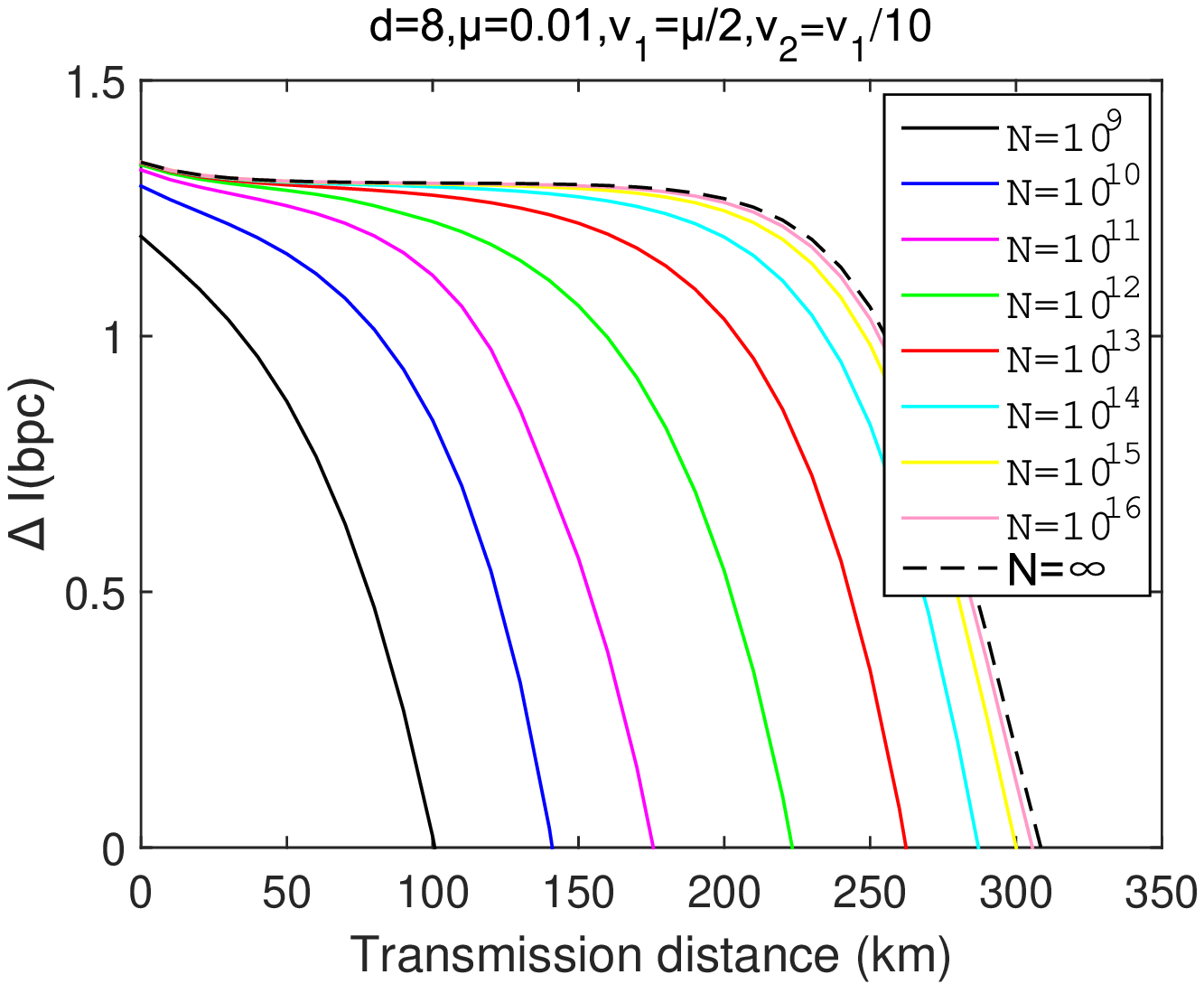}}
  \subfigure[]{\label{fig:3-2}\includegraphics[width=2.0in]{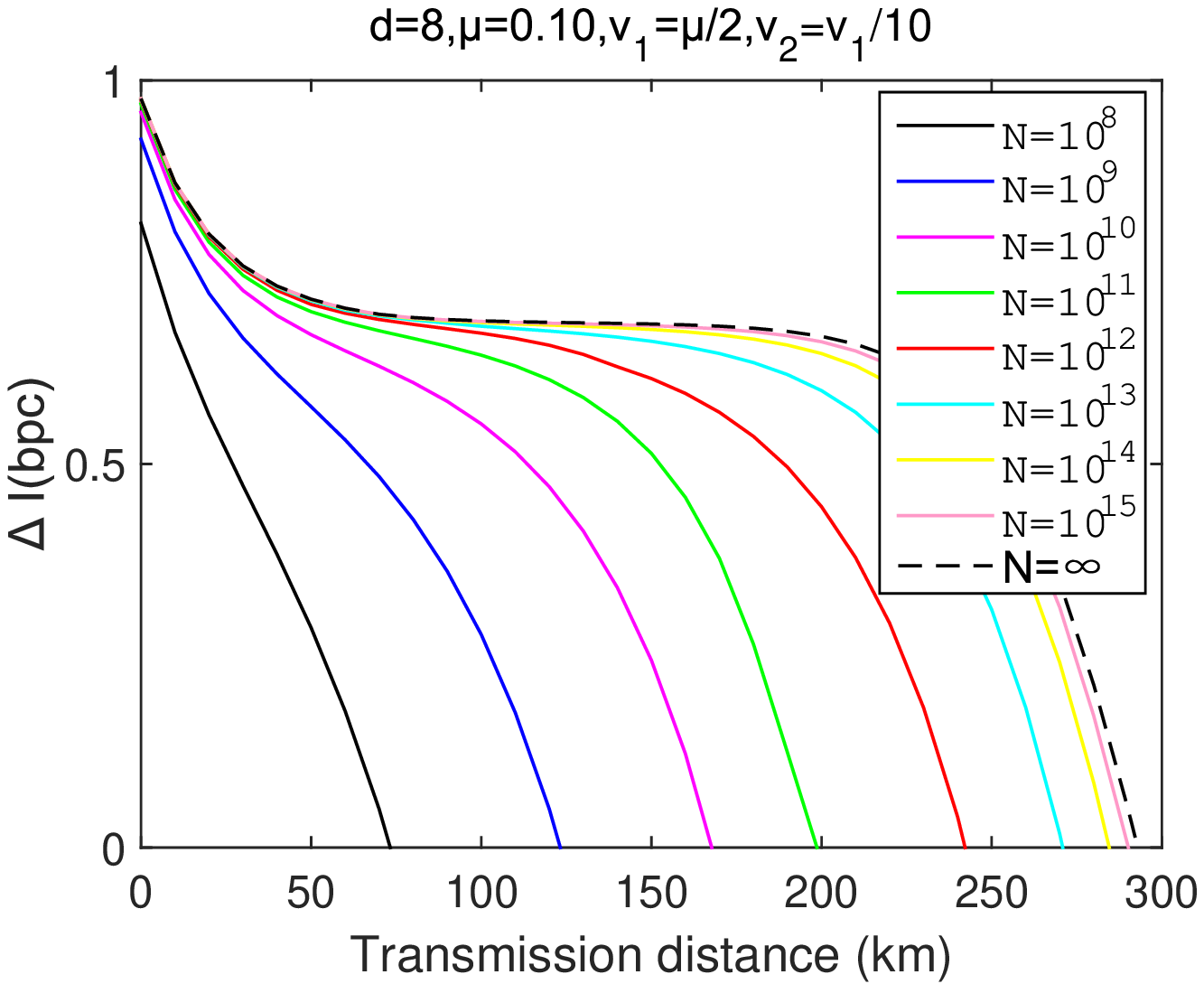}}
  \subfigure[]{\label{fig:3-3}\includegraphics[width=2.0in]{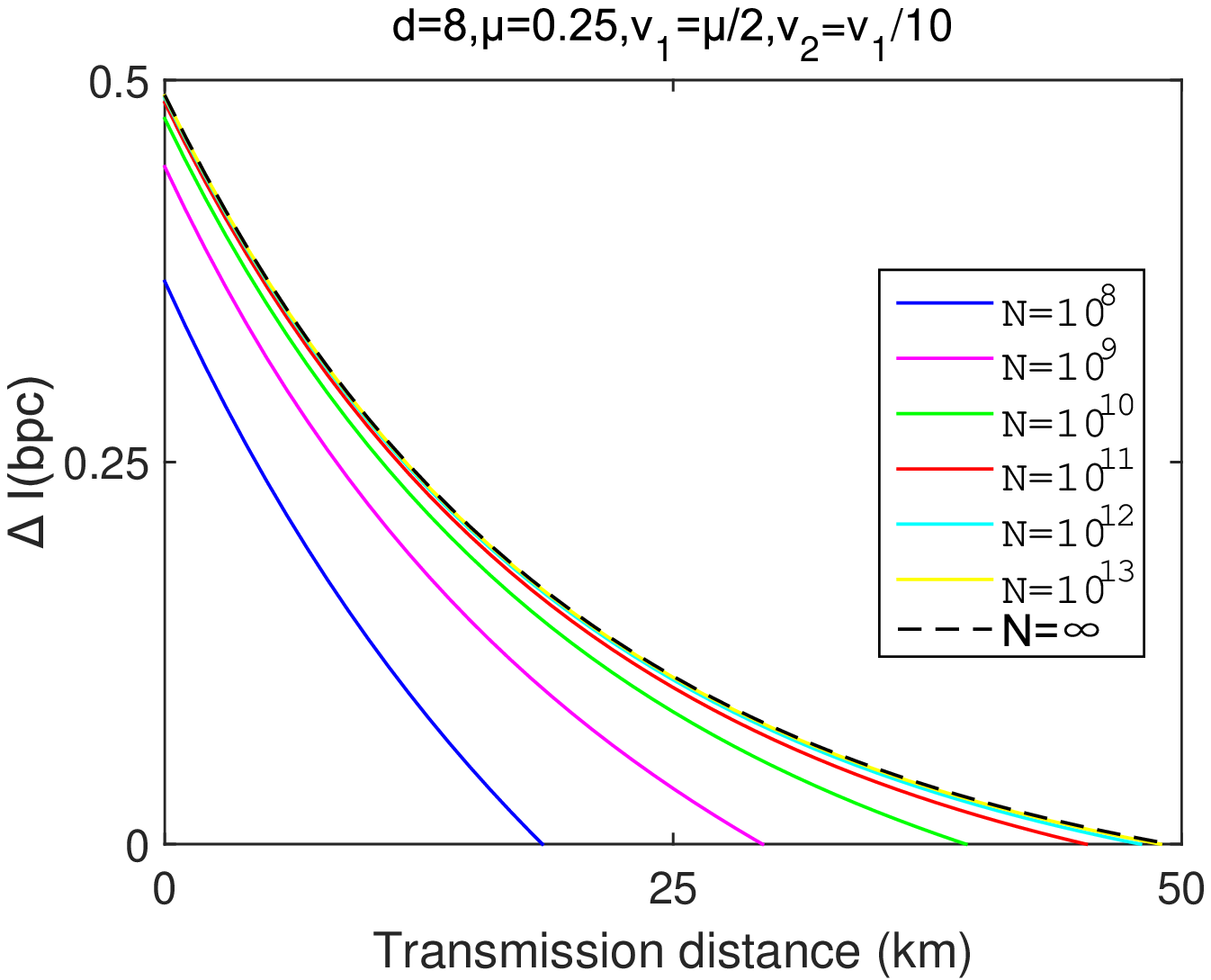}}
  \\
  \subfigure[]{\label{fig:3-4}\includegraphics[width=2.0in]{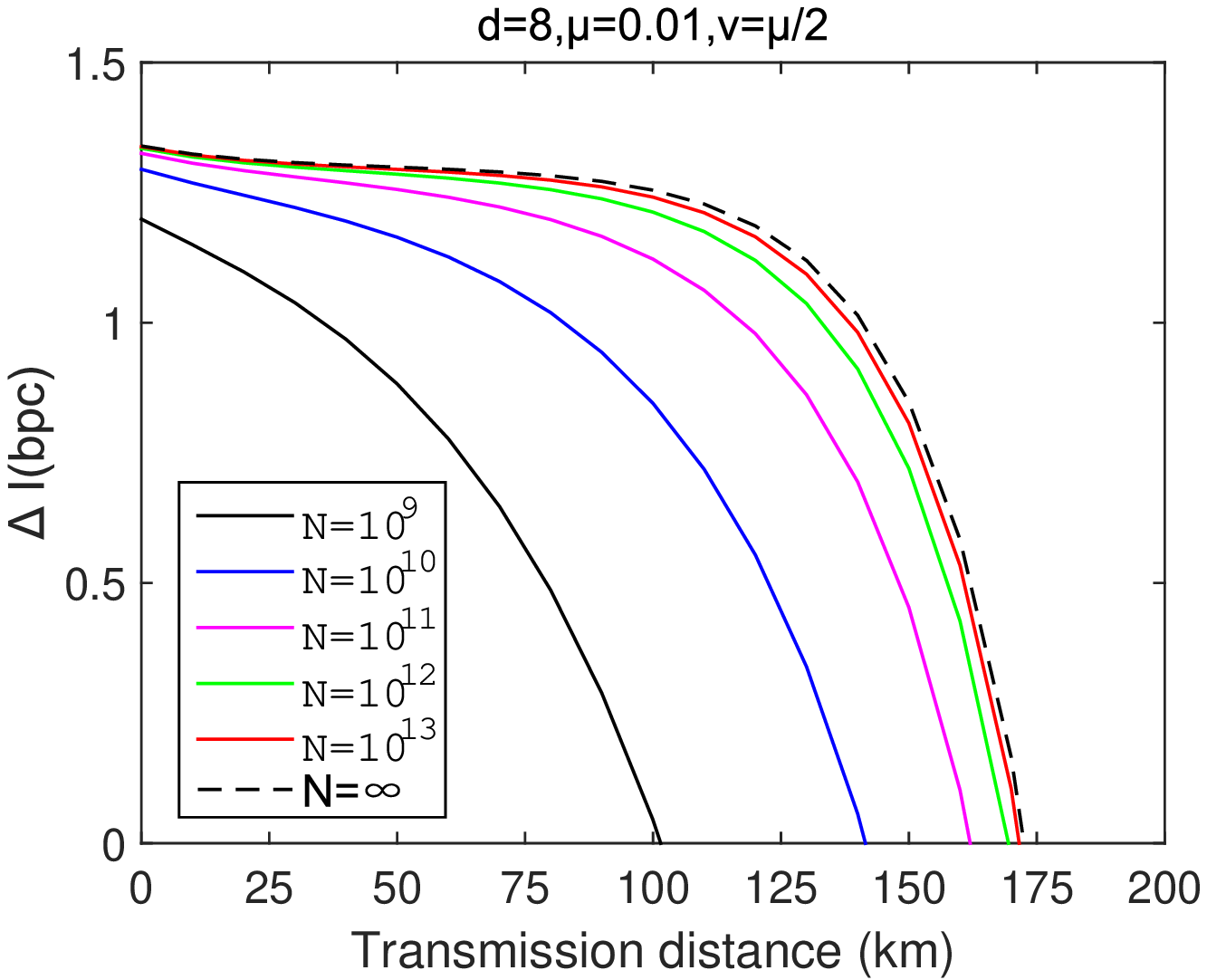}}
  \subfigure[]{\label{fig:3-5}\includegraphics[width=2.0in]{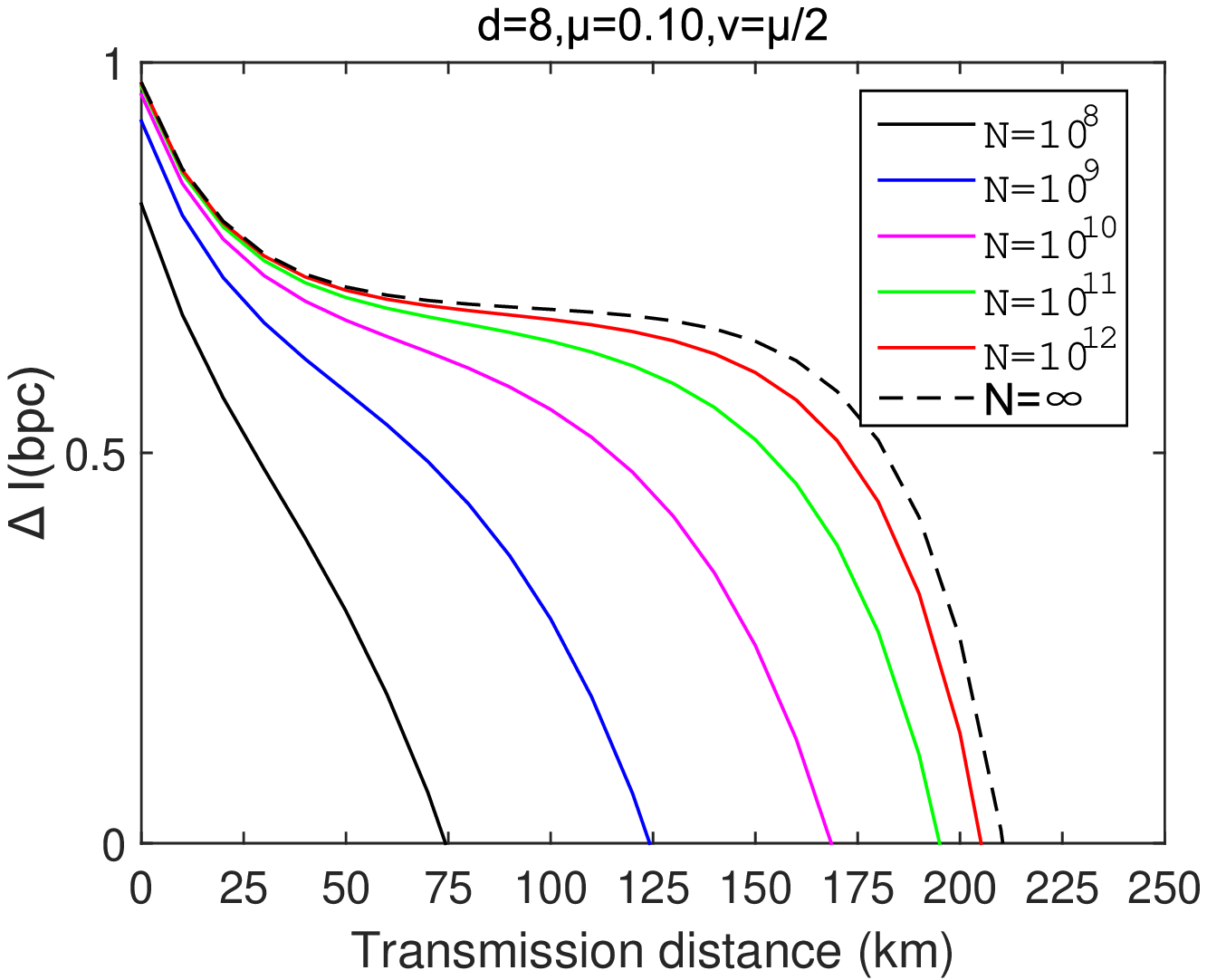}}
  \subfigure[]{\label{fig:3-6}\includegraphics[width=2.0in]{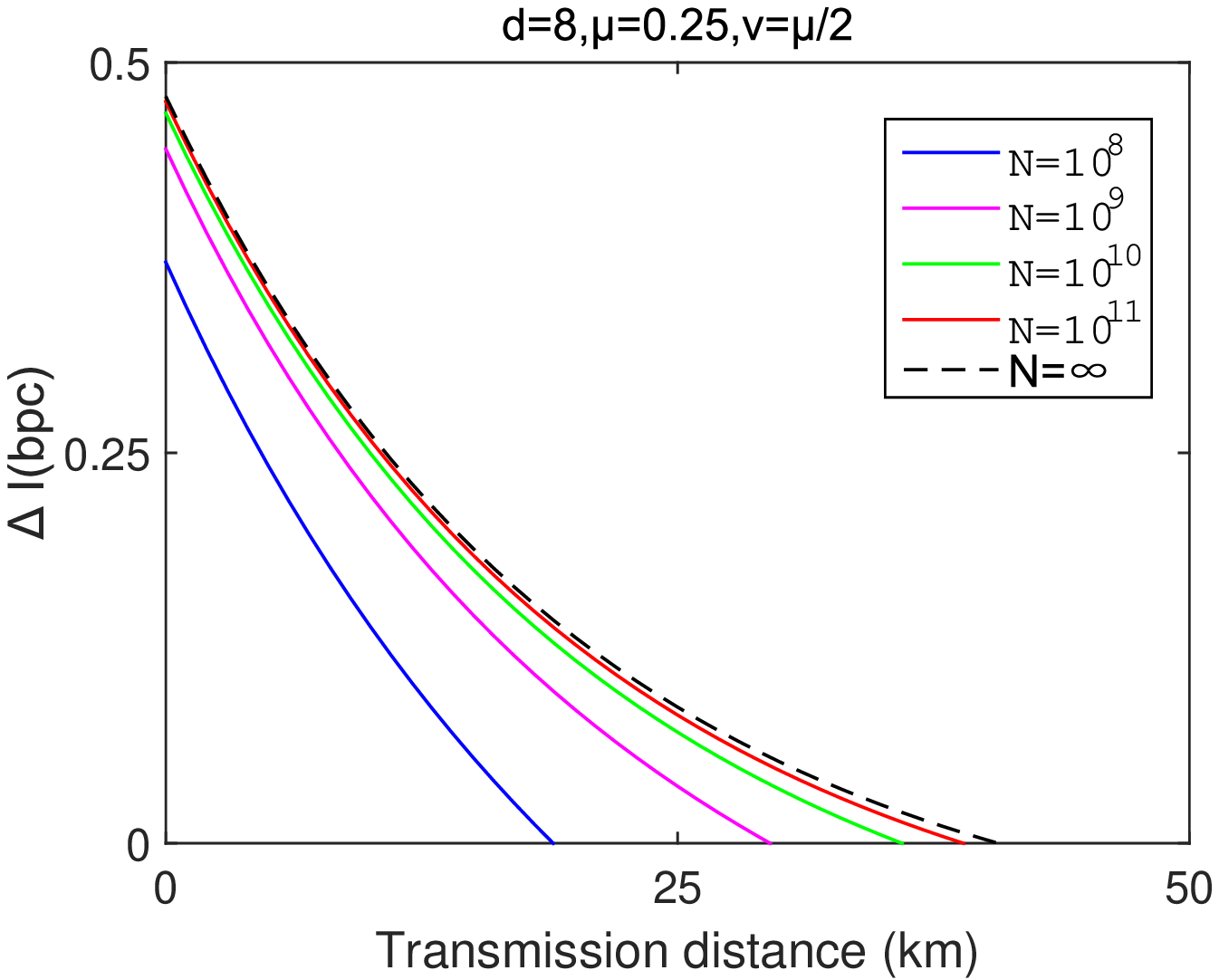}}
  \caption{(Color online) Finite-size secure-key capacities in bpc versus transmission distance based on Chernoff bound. }
  \label{fig3}
\end{figure}

In decoy-state QKD scenarios, the intensity of decoy states is far weaker than signal states \cite{21,40,41,42}. To maximize the security bound in any particular transmission distance, in the original decoy-state HD-QKD protocol, ${{v}_{2}}$, the intensity of the weaker decoy state, is optimized. But in fact, when ${{v}_{2}}$ is weak enough and fixed, we can't obviously distinguish the security bound with the optimized security bound. For simplifying evaluation, we fix ${{v}_{2}}$ at ${{v}_{2}}={{{v}_{1}}}/{10}\;$, which is reasonable and doesn't lose generality.

According to Fig~\ref{fig2} and Fig~\ref{fig3}, we can see that in the infinite-key regime, the two-decoy-state HD-QKD protocol can achieve a longer transmission distance than the single-decoy-state HD-QKD protocol under the same condition. We denote the maximum transmission distance of the two-decoy-state protocol as $L _{N}^{\left\{{{{v }_{1}},{{v }_{2}}}\right\}}$ and the maximum transmission distance of the single-decoy-state as $L _{N}^{\left\{{v}\right\}}$. Here $L _{N=\infty}^{\left\{{{{v }_{1}},{{v }_{2}}}\right\}}$ is the maximum transmission of two-decoy-state HD-QKD protocol and $L _{N=\infty}^{\left\{{v}\right\}}$ is the maximum transmission of single-decoy-state HD-QKD protocol in the infinite-key regime.

Comparing Subfigure~\ref{fig:2-1} with ~\ref{fig:2-4}, ~\ref{fig:2-2} with ~\ref{fig:2-5}, ~\ref{fig:2-3} with ~\ref{fig:2-6} in Figure~\ref{fig2} and Subfigure~\ref{fig:3-1} with ~\ref{fig:3-4}, ~\ref{fig:3-2} with ~\ref{fig:3-5}, ~\ref{fig:3-3} with ~\ref{fig:3-6} in Figure~\ref{fig3}, we find that when the number of signals $N$ is small, the single-decoy-state protocol can achieve similar results to the two-decoy-state protocol, i.e. in Figure~\ref{fig:3-1} and Figure~\ref{fig:3-4}, the secure-key capacity and the maximum transmission distance are almost the same when $N=10^{9}, \ 10^{10}$ and $10^{11}$. With $N$ increasing, $L _{N}^{\left\{{v}\right\}}$ ($L _{N}^{\left\{{{{v }_{1}},{{v }_{2}}}\right\}}$) gets close to $L _{N=\infty}^{\left\{{v}\right\}}$ ($L _{N=\infty}^{\left\{{{{v }_{1}},{{v }_{2}}}\right\}}$). When $N$ gets big enough, the results achieved in Figures demonstrate that the two-decoy-state HD-QKD protocol has an obvious advantage over the single-decoy-state protocol. For example, in Figure~\ref{fig:3-1} and Figure~\ref{fig:3-4}, when $N=10^{11}$ , the results of two figures are still similar, but when $N=10^{12}$, $L _{N}^{\left\{{{{v }_{1}},{{v }_{2}}}\right\}}$ is much greater than $L _{N}^{\left\{{v}\right\}}$.

We note that when $N$ is fixed under the same condition, with the transmission distance extending, the secure-key capacity estimated by Hoeffding's inequality descends faster in Figure~\ref{fig2} than that estimated by Chernoff bound in Figure~\ref{fig3}. So we can get a better effect in parameter estimation by using Chernoff bound than using Hoeffding's inequality in the finite-key decoy-state HD-QKD protocol. In particular, we consider the single-decoy-state protocol in the case of $d=8$, $\mu=0.10$ and $d=8$, $\mu=0.25$. The result is shown in Figure~\ref{4}. We can see that in a short transmission distance, the result estimated by Chernoff bound is not better than Hoeffding's inequality. With the transmission distance extending, the advantage of Chernoff bound gets more and more obvious.
\begin{figure}
  \centering
  \subfigure[]{
    \label{fig:4-1}
    \includegraphics[width=2.6in]{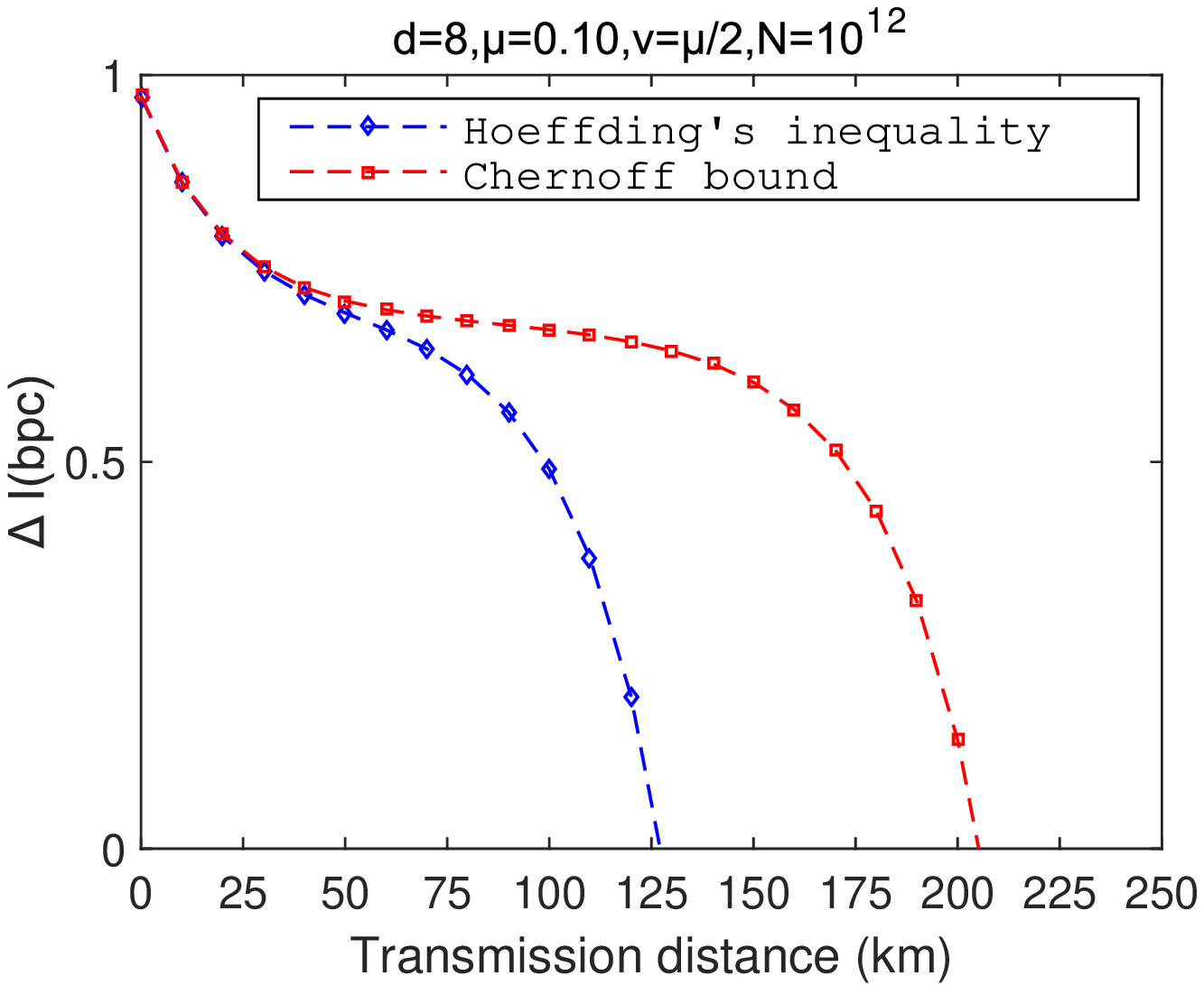}}
  \hspace{0in}
  \subfigure[]{
    \label{fig:4-2}
    \includegraphics[width=2.6in]{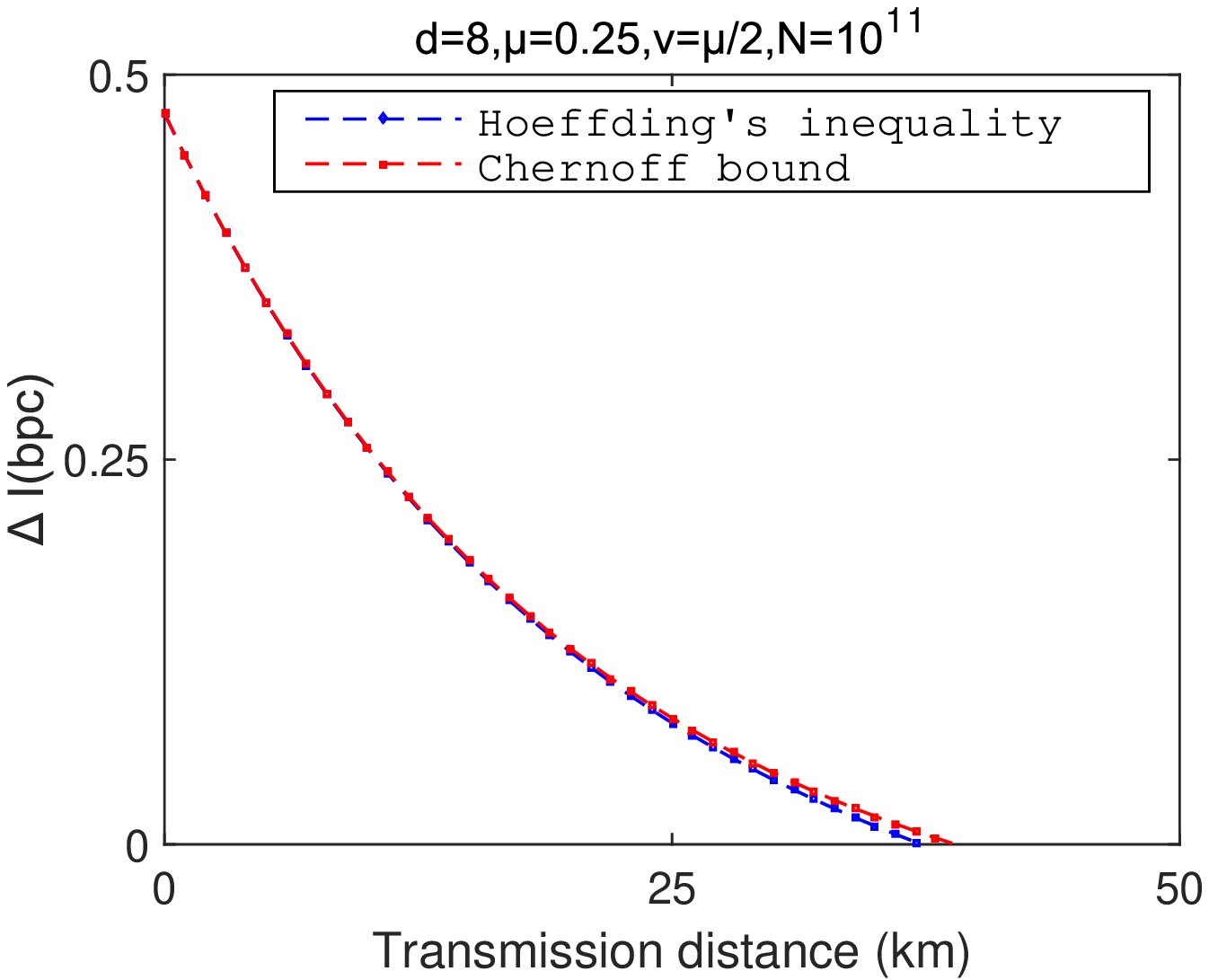}}
  \caption{(Color online) (4-1) Chernoff bound and Hoeffding's inequality in the single-decoy-state protocol with the case $d=8$, $\mu =0.10$ and $N{{=10}^{12}}$. (4-2) Chernoff bound and Hoeffding's inequality in the single-decoy-state protocol with the case $d=8$, $\mu =0.25$ and $N{{=10}^{11}}$.}
  \label{4}
\end{figure}

Ref.\cite{23} has indicated that when the dimensionality is increased in the infinite-key regime, the secure-key capacity would be promoted. Besides, if the intensity of source is strong enough, the maximum transmission distance would be extended significantly. To clarify the effect of the dimensionality under the constraint of finite keys, Figure~\ref{fig5} plots the secure-key capacity with Schmidt number $d=32$ and $\mu =0.01$ in the single-decoy-state HD-QKD protocol and Figure~\ref{6} plots the secure-key capacity with Schmidt number $d=32$ and $\mu =0.10$ and 0.25 in the two-decoy-state HD-QKD protocol.
\begin{wrapfigure}{l}{0.5\textwidth}
  \centering
  \includegraphics[width=2.6in]{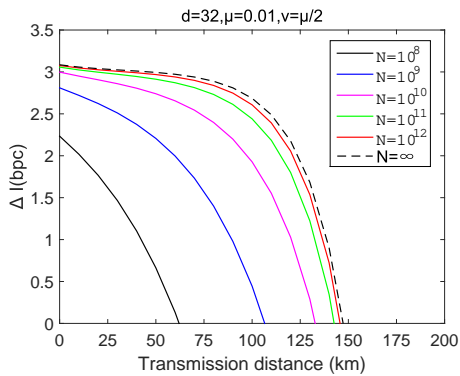}
  \caption{(Color online) Plot the secure-key capacity of the single-decoy-state HD-QKD protocol with the Schmidt number $d=32$ and $\mu =0.01$.}
  \label{fig5}
\end{wrapfigure}

In two-decoy-state HD-QKD protocol, by comparing Figure~\ref{fig:3-4} with Figure~\ref{fig5} ,Figure~\ref{fig:3-2} with Figure~\ref{fig:6-1}, we can find that when $N$ is fixed, the 32-dimension QKD protocol can achieve further transmission distance than 8-dimension QKD protocol. If the intensity of the source is strong enough, using fewer signals in 32-dimension QKD protocol can reach or even transcend the result that achieved by more signals in 8-dimension QKD protocol. For example, in 32-dimension QKD protocol with $\mu =0.25$, just $10^{8}$ signals could achieve the better result than that achieved in infinite-key regime in 8-dimension QKD protocol. The details are shown in Figure~\ref{fig:3-3} and Figure~\ref{fig:6-2}.
\begin{figure}
  \centering
  \subfigure[]{
  \label{fig:6-1}
    \includegraphics[width=2.6in]{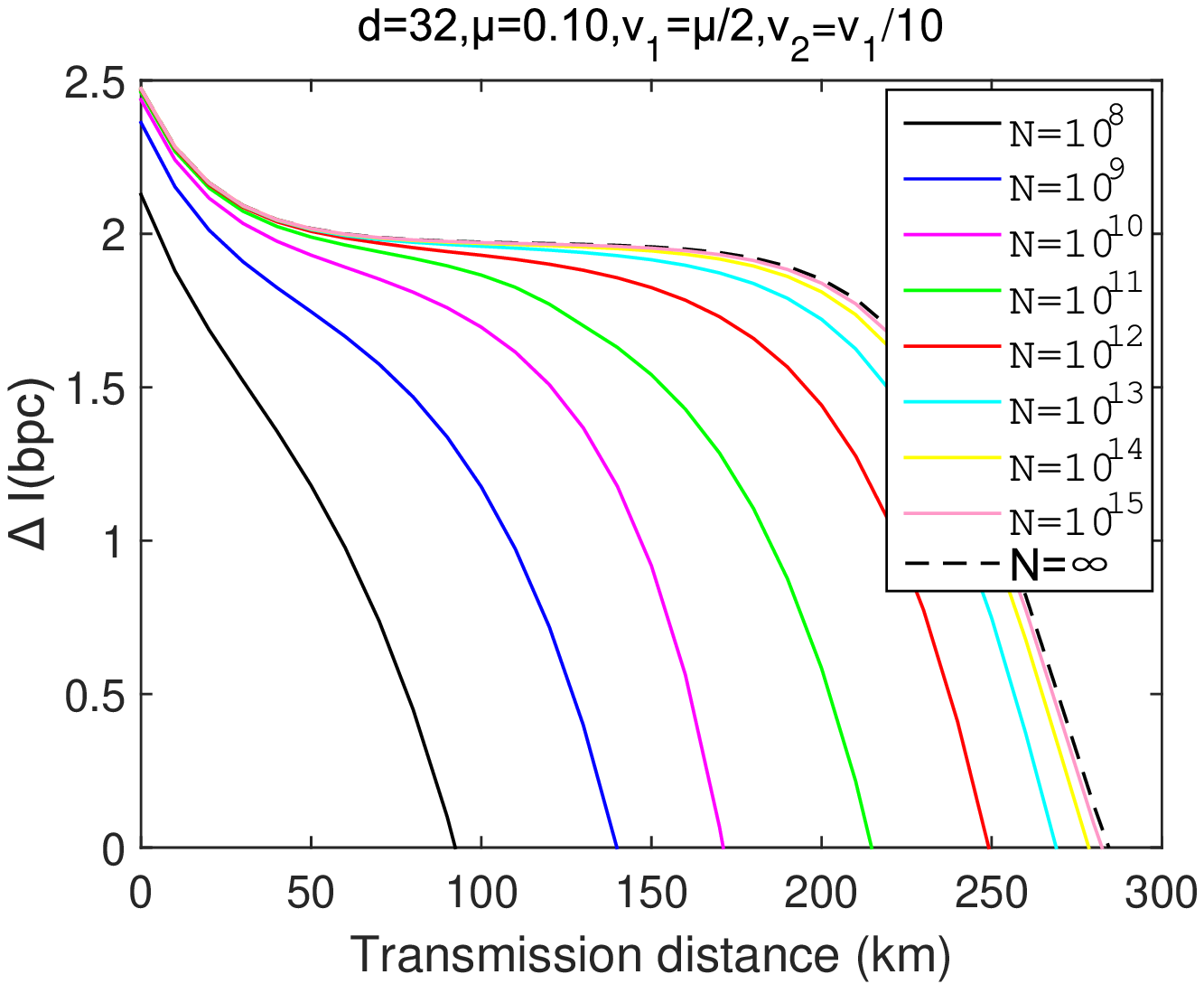}}
  \subfigure[]{
  \label{fig:6-2}
    \includegraphics[width=2.6in]{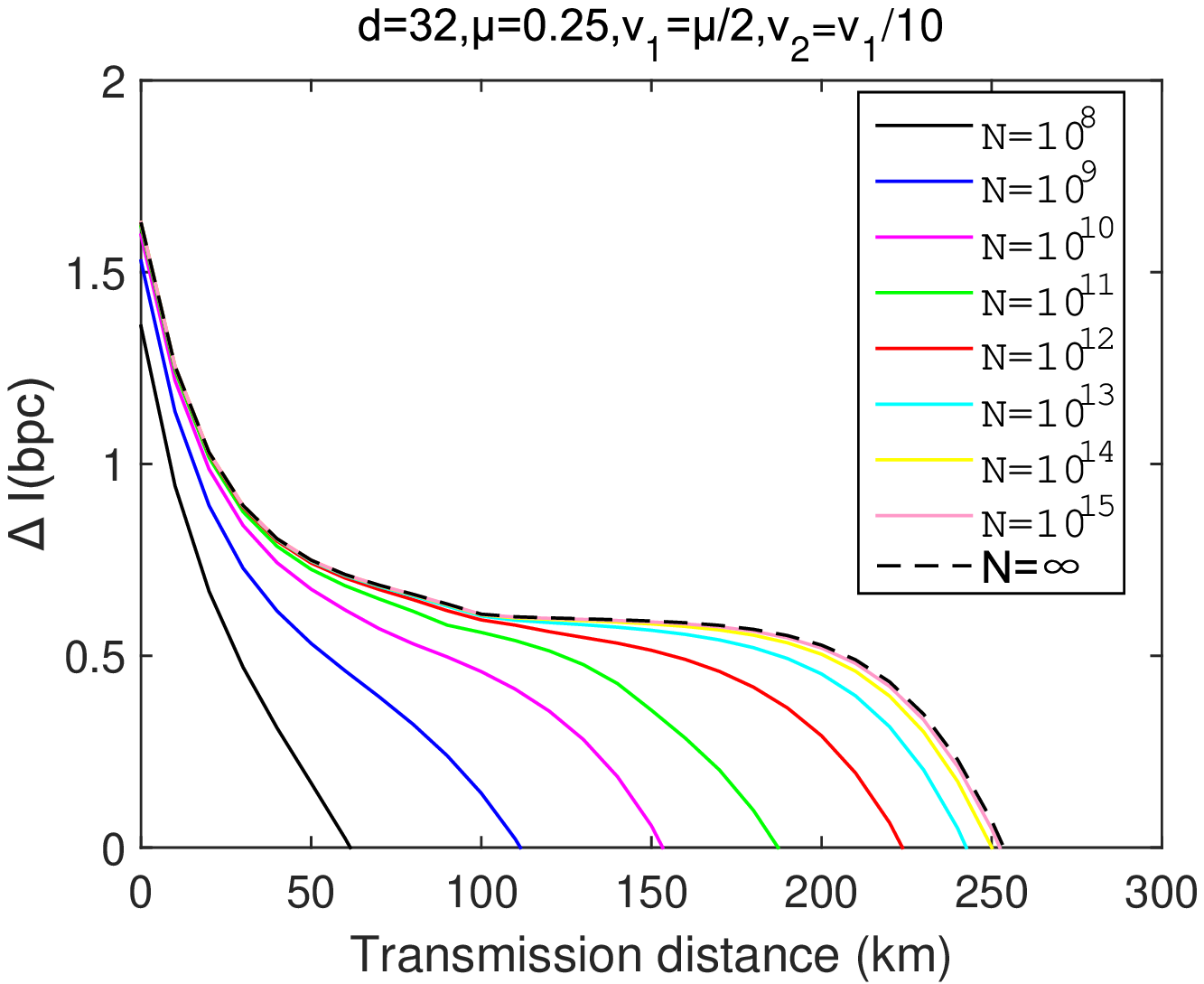}}
  \caption{(Color online) Plot the secure-key capacity of the two-decoy-state HD-QKD protocol with the Schmidt number $d=32$ and $\mu =0.10$ and $0.25$.}
  \label{6}
\end{figure}

In realistic condition, when the transmission distance is settled, for increasing the secure-key capacity and meanwhile maximizing the secure-key capacity with minimum pulses, two authorized parties can choose an appropriate protocol and optimized parameters such as source intensities, channel noise and probabilities of choosing measurement basis and ratios between signal states and decoy states.

\section{Conclusions}
We have used two methods to analyze the security against Gaussian collective attacks of the HD-QKD protocol in the finite-key regime. By numerical evaluation, we have shown that in a long transmission distance, more precise results could be achieved in parameter estimation by using Chernoff bound. If the number of the signal pulses is small, the single-decoy-state HD-QKD protocol could reach similar results to the two-decoy-state protocol. Some assumptions are still needed to be noted. In particular, we require that the detection probability of Alice's and Bob's measurement devices is independent of their basis choice.

We just considered the case of HD-QKD with two decoy states and with one decoy state, which is based on dispersive optics, but the arguments presented in this work can be generalized to the case of multi-decoy-state HD-QKD protocol and other HD-QKD protocols based on other degrees of freedom. When an experiment based on the HD-QKD with decoy states is implemented practically, some parameters such as source intensities, channel noise and probabilities in choosing measurement basis and ratios between signal states and decoy states could be optimized so as to maximize secure-key capacities. In realistic constraints, the method presented in this paper provides a theoretical approach to optimized parameters so as to maximize the rate with minimum resources.

Besides our methods, many other techniques, such as the random sampling technique\cite{Fung}, are used in analyzing finite-size effect. So our future work will focus on whether a tighter security bound could be obtained by other techniques, which is worthy and necessary.
\ack
This work was supported by the National Basic Research Program of China (Grant No. 2013CB338002) and the National Natural Science Foundation of China (Grants No. 11304397 and No. 61505261).

\end{document}